\documentclass[fleqn,usenatbib]{mnras}

\usepackage{newtxtext,newtxmath}

\usepackage[T1]{fontenc}

\DeclareRobustCommand{\VAN}[3]{#2}
\let\VANthebibliography\thebibliography
\def\thebibliography{\DeclareRobustCommand{\VAN}[3]{##3}\VANthebibliography}

\usepackage{graphicx}	
\usepackage{amsmath}	

\usepackage{float} 

\def\kms{km\,s$^{-1}$}
\graphicspath{{Figures/}}
\usepackage[dvipsnames]{xcolor}
\usepackage{orcidlink}
\usepackage{longtable}
\usepackage{multicol}

\title[SN2019szu - PPI Candidate]{A Precursor Plateau and Pre-Maximum [O II] Emission in the Superluminous SN2019szu: A Pulsational Pair-Instability Candidate}

\author[A. Aamer et al.]{Aysha Aamer$^{1,2}$\thanks{E-mail: aa@star.sr.bham.ac.uk (AA)}\orcidlink{0000-0002-9085-8187},
Matt Nicholl$^{2}$\orcidlink{0000-0002-2555-3192},
Anders Jerkstrand$^{3}$\orcidlink{0000-0001-8005-4030},
Sebastian Gomez$^{4}$\orcidlink{0000-0001-6395-6702},
Samantha R. Oates$^{1}$,
\newauthor
Stephen J. Smartt$^{2,5}$\orcidlink{0000-0002-8229-1731},
Shubham Srivastav$^{2}$\orcidlink{0000-0003-4524-6883},
Giorgos Leloudas$^{6}$,
Joseph P. Anderson$^{7,8}$\orcidlink{0000-0003-0227-3451},
Edo Berger$^{9}$,
\newauthor
Thomas de Boer$^{10}$,
Kenneth Chambers$^{10}$\orcidlink{0000-0001-6965-7789},
Ting-Wan Chen$^{11}$\orcidlink{0000-0002-1066-6098},
Llu\'is Galbany$^{12,13}$\orcidlink{0000-0002-1296-6887},
Hua Gao$^{10}$\orcidlink{0000-0003-1015-5367},
\newauthor
Benjamin P. Gompertz$^{1}$,
Maider Gonz\'alez-Ba\~nuelos$^{12,14}$,
Mariusz Gromadzki$^{15}$\orcidlink{0000-0002-1650-1518},
Claudia P. Guti\'errez$^{13,12}$,
\newauthor
Cosimo Inserra$^{16}$\orcidlink{0000-0002-3968-4409},
Thomas B. Lowe$^{10}$,
Eugene A. Magnier$^{10}$\orcidlink{0000-0002-7965-2815},
Paolo A. Mazzali$^{12,17}$,
Thomas Moore$^{2}\orcidlink{0000-0001-8385-3727}$,
\newauthor
Tom\'as E. M\"uller-Bravo$^{12,13}$\orcidlink{0000-0003-3939-7167},
Miika Pursiainen$^{6}$,
Armin Rest$^{4,19}$\orcidlink{0000-0002-4410-5387},
Steve Schulze$^{3}$,
Ken W. Smith$^{2}$\orcidlink{0000-0001-9535-3199},
\newauthor
Jacco H. Terwel$^{20,21}$\orcidlink{0000-0001-9834-3439},
Richard Wainscoat$^{10}$\orcidlink{0000-0002-1341-0952},
David R. Young$^{2}$\orcidlink{0000-0002-1229-2499}
\\
\\
\textit{Affiliations are listed at the end of the paper}}

\date{Accepted XXX. Received YYY; in original form ZZZ}

\pubyear{2023}

\begin{document}
\label{firstpage}
\pagerange{\pageref{firstpage}--\pageref{lastpage}}
\maketitle

\begin{abstract}

We present a detailed study on SN2019szu, a Type I superluminous supernova at $z=0.213$, that displayed unique photometric and spectroscopic properties. Pan-STARRS and ZTF forced photometry shows a pre-explosion plateau lasting $\sim$ 40 days. Unlike other SLSNe that show decreasing photospheric temperatures with time, the optical colours show an apparent temperature increase from $\sim$15000\,K to $\sim$20000\,K over the first 70 days, likely caused by an additional pseudo-continuum in the spectrum. Remarkably, the spectrum displays a forbidden emission line (likely attributed to $\lambda\lambda$7320,7330) visible 16 days before maximum light, inconsistent with an apparently compact photosphere. This identification is further strengthened by the appearances of [O III] $\lambda\lambda$4959, 5007, and [O III] $\lambda$4363 seen in the spectrum. Comparing with nebular spectral models, we find that the oxygen line fluxes and ratios can be reproduced with $\sim$0.25\,M$_{\odot}$ of oxygen rich material with a density of $\sim10^{-15}\,\rm{g\,cm}^{-3}$. The low density suggests a circumstellar origin, but the early onset of the emission lines requires that this material was ejected within the final months before the terminal explosion, consistent with the timing of the precursor plateau. Interaction with denser material closer to the explosion likely produced the pseudo-continuum bluewards of $\sim$5500\,$\Angstrom$. We suggest that this event is one of the best candidates to date for a pulsational pair-instability ejection, with early pulses providing the low density material needed for the formation of the forbidden emission line, and collisions between the final shells of ejected material producing the pre-explosion plateau.

\end{abstract}




\section{Introduction}
\label{sec:intro}
Superluminous supernovae (SLSNe) are a class of supernovae (SNe), initially categorised as events with absolute magnitudes exceeding $M<-21$ mag \citep{Quimby2011, Gal-Yam2012}.  In addition to their high luminosities, these events radiate $\sim10^{51}$\,erg when integrated over their broad light curves \citep{Gal-Yam2012,Lunnan2018b}. The luminous nature of these events means we can observe them out to redshifts $z>4$ \citep{Cooke2012}, and so even though the volumetric rate of these events is $\sim 1$ in a few thousand SNe \citep{Quimby2013, Frohmaier2021}, they make up roughly 1$\%$ of the SNe discovered today \citep{Fremling2020}. This is aided by the wide-field surveys available which can probe the entire night sky instead of targeting only nearby massive galaxies. SLSNe were missed by earlier surveys due to their preference for metal-poor dwarf galaxies as hosts \citep{Chen2013, Lunnan2014, Leloudas2015b, Perley2016, Schulze2018}. As more of these events have been discovered, the strict magnitude cut off for classification has since been replaced by spectral classification around peak luminosity \citep{Gal-Yam2019b, Quimby2018}, driven by events with SLSN-like spectra but intermediate luminosities \citep{DeCia2018, Angus2019, Gomez2022}. 

SLSNe can further be classified into Type I and Type II SLSNe, analogous to their less luminous counterparts. Type II SLSNe often resemble lower luminosity SNe IIn, with narrow hydrogen lines and a small subset displaying broad hydrogen lines \citep{Kangas2022}. Interaction with circumstellar material (CSM) is thought to the main power source for Type II SLSNe \citep{Ofek2014, Inserra2018}. Type I SLSNe (often simply called SLSNe) lack hydrogen in their spectra. These spectra are characterised by a steep blue continuum indicative of high temperatures, and often show prominent O II absorption lines at early times \citep{Quimby2011}, eventually evolving to be similar to SNe Ic when at comparable temperatures \citep{Pastorello2010}.  However, a small fraction of these events, show evidence of H$\alpha$ at late times \citep[$\gtrsim 30$ days;][]{Pursiainen2022,Yan2018} but this is not necessarily related directly to the power source at maximum light, and instead may be a product of interaction with its environment at scales $\gtrsim 10^{16}$\,cm \citep{Yan2015}. A handful of events have also shown evidence for helium in their photospheric spectra \citep{Quimby2018, Yan2020}.

One of the biggest questions remaining about SLSNe pertains to their powering mechanism. Typical hydrogen-poor SNe are powered by the decay of radioactive nickel ($^{56}$Ni), but this explanation does not seem to work for most SLSNe for a number of reasons. Powering the peak luminosities of SLSNe with nickel decay would require $\sim 5-20$\,M$_\odot$ of $^{56}$Ni, which is too high compared to the inferred ejected mass from the light curves \citep{Inserra2013b, Blanchard2018}. This amount of nickel could likely only be produced in Pair-Instability SNe (PISN) of stars with initial masses $M \gtrsim 140$\,M$_{\odot}$ \citep{Heger2002, Gal-Yam2009}. One of the best PISN candidates to date SN2018ibb is thought to be powered by $25-44$\,M$_\odot$ of freshly synthesised $^{56}$Ni produced by a star with a helium core mass of $120-130$\,M$_\odot$ \citep{Schulze2023}. Although stars in this mass range required for PISNe have been observed, mass-loss on the main sequence makes PISN formation challenging except at very low metallicities \citep{Yusof2013}. However, some models suggest a magnetic field at the surface of the star could quench the mass loss for stars at solar metallicity, allowing enough mass to remain for the PI mechanism \citep{Georgy2017}. More problematic for SLSNe, any model producing this amount of nickel would result in a spectrum dominated by iron-group elements, which is inconsistent with the blue spectra of SLSNe \citep{Dessart2012, Nicholl2013, Jerkstrand2016}. However, these models often assume interaction between the SN ejecta and CSM is negligible which would only likely be the case if observed early enough that explosive nucleosynthesis is not affecting the layers of ejecta \citep{Kasen2011}.

Some theories suggest an internal power source such as the spin down energy of a magnetar or an accreting black hole could power SLSNe. However the accreted mass required in the latter scenario ($\gg 100$\,M$_\odot$) often exceeds the mass of any reasonable star \citep{Moriya2018}. In the magnetar scenario, the remnant is a fast rotating neutron star with a very strong magnetic field $B \sim 10^{13}-10^{14}$ G \citep{Kasen2010}. Nearly $10\%$ of newly born neutron stars have B-fields in the range $10^{13}-10^{15}$ G lasting over 1000 years after their birth, and so it is plausible that these could exist to produce SLSNe  \citep{Kouveliotou1998, Woods2006}. This mechanism could explain the long duration of the light curves as the magnetar releases its rotational energy at the dipole spin-down rate, which remains high for days to weeks in this range of magnetic field strengths \citep{Ostriker1971, Kasen2010}. 

Another possible explanation to power SLSNe is interaction with CSM. This theory proposes a core collapse supernova that has large amounts of CSM created through stellar winds and ejections throughout the life of the progenitor star. The SN ejecta is able to catch up to this material because it has much higher velocities, and is rapidly decelerated if the CSM is massive enough. This creates a shock that deposits energy in the ejecta and CSM, the cooling of which can produce a bright and long-lived light curve \citep{Chevalier2011, Smith2007b}. The mass of CSM required to efficiently power a bright light curve must be comparable to the ejecta mass \citep{Chevalier2011, Ginzburg2012}, ranging from a few solar masses up to a few tens \citep{Nicholl2014}. Getting this much mass close to the star just before explosion is difficult to explain using stellar winds, even in Wolf Rayet stars with high mass loss rates \citep{Mauron2011, Sander2022}. An alternative way to produce massive CSM is through discrete outbursts. Stars with masses in the range $70-140$\,M$_{\odot}$ are thought to undergo pulsational pair instability PPI eruptions \citep{Woosley2017}. These stars are not massive enough to experience terminal pair instability, instead the star violently expels up to tens of solar masses worth of material towards the end of its life due to this mechanism \citep{Woosley2017}. This has been suggested as a way to get sufficiently massive CSM to power SLSNe \citep{Woosley2007}.

The CSM model has been questioned as the main power source for hydrogen-poor SLSNe. We would naively expect to see narrow lines in the spectra from slow moving material if interaction was at play, but this is not seen in all SLSNe \citep{Nicholl2015a}. Instead, there is evidence that CSM interaction may play a role in powering some SLSNe including late time interaction producing H$\alpha$ emission \citep{Yan2018, Pursiainen2022}, post peak bumps in the light curves \citep{Hosseinzadeh2022}, blue pseudo-continua, and early appearances of forbidden emission lines such as [O II] and [O III] \citep{Schulze2023}. Light curve and spectral modelling suggest that both central engines and CSM interaction may help to power this class of events \citep{Chen2017, Chen2023b}.

Spectra taken at different phases of the SN evolution allow us to probe different regions of the ejecta. At early times the ejected material from the explosion is still optically thick and obscures the view of the inner layers. As the ejecta expands it becomes less dense, leading to more states leaving local thermal equilibrium (LTE) and lower populations of excited states, reducing the number of optically thick lines and bound-free continua. This spectral transition from photospheric to nebular is also driven by decreasing temperatures which results in fewer lines that are capable of significant cooling, and fewer excited states with enough population to provide opacity. This transition occurs typically on the timescale of hundreds of days and results in a spectrum dominated by low-lying forbidden transitions \citep{Jerkstrand2017a}. This contradicts observations in which some SLSNe have shown these forbidden emission lines early on during their photospheric phase. This includes SN2018ibb which displayed signs of a possible [Ca II] $\lambda\lambda$7291,7323 at -1.4 days before peak, becoming prominent by 30 days later \citep{Schulze2023}. Other SLSNe have also shown signs of early forbidden emission lines including the earliest spectra of SN2007bi \citep{Gal-Yam2009}, and LSQ14an \citep{Inserra2017}, in which this line $\sim50$ days after peak was also attributed to [Ca II]. This suggests some lower density regions exist in the ejecta or their surroundings. In principle these lines could have appeared even earlier if earlier spectra were obtained, challenging our understanding of the structure of this massive ejecta.

In this paper we present and analyse SN2019szu, a slowly evolving SLSN that showed forbidden emission lines remarkably soon after the time of explosion, at least 16 days before maximum light. We identify these lines as singly- and doubly-ionized oxygen, arising in a low density, hydrogen-poor CSM, and use this to place important constraints on the progenitor of this event. The structure of this paper is as follows. Section \ref{sec:observations} outlines the data collected for this object. Section \ref{sec:analysis} covers the analysis of the host galaxy, and the photometric and spectroscopic data collected for the target. In Section \ref{sec:Models}, we discuss spectral models to fit this event as well as \textsc{mosfit} models of the light curve. We then discuss the implications of these results and how they fit into our understanding of SN2019szu in Section \ref{sec:discussion}. Lastly, in Section \ref{sec:conclusion} we present our conclusions.

\begin{figure*}
	\includegraphics[width=2\columnwidth]{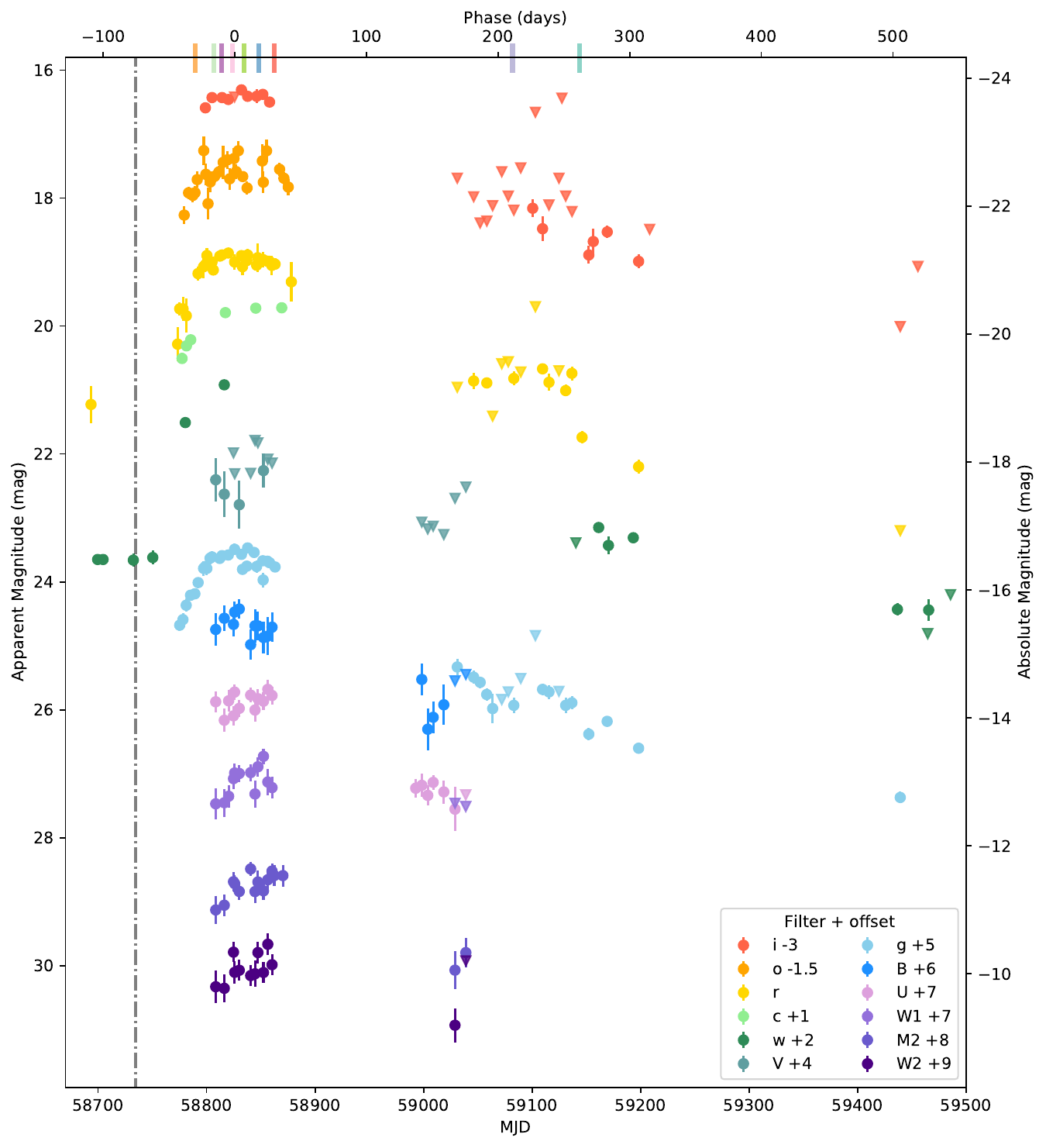}
    \caption{Light curve of SN2019szu. All magnitudes are in given in the AB system and are not corrected for Milky Way extinction. The grey vertical line indicates time of explosion based on light curve fits from \textsc{mosfit}. Phase given in rest frame days with respect to maximum light in the $g$-band. Includes data from ATLAS, Pan-STARRS, ZTF, \textit{Swift}, LCO, and NTT. MJD is in the observer frame and $3\sigma$ upper limits are indicated via inverted triangles. The ATLAS $o$ and $c$ bands are plotted without upper limits for clarity and the o band data points are binned to a 2 day cadence. Vertical markers on the top axis correspond to the spectra in Figure \ref{fig:spec ev} and indicate when they were obtained.}
    \label{fig:light curve split}
\end{figure*}

\begin{figure*}
	\includegraphics[width=2\columnwidth]{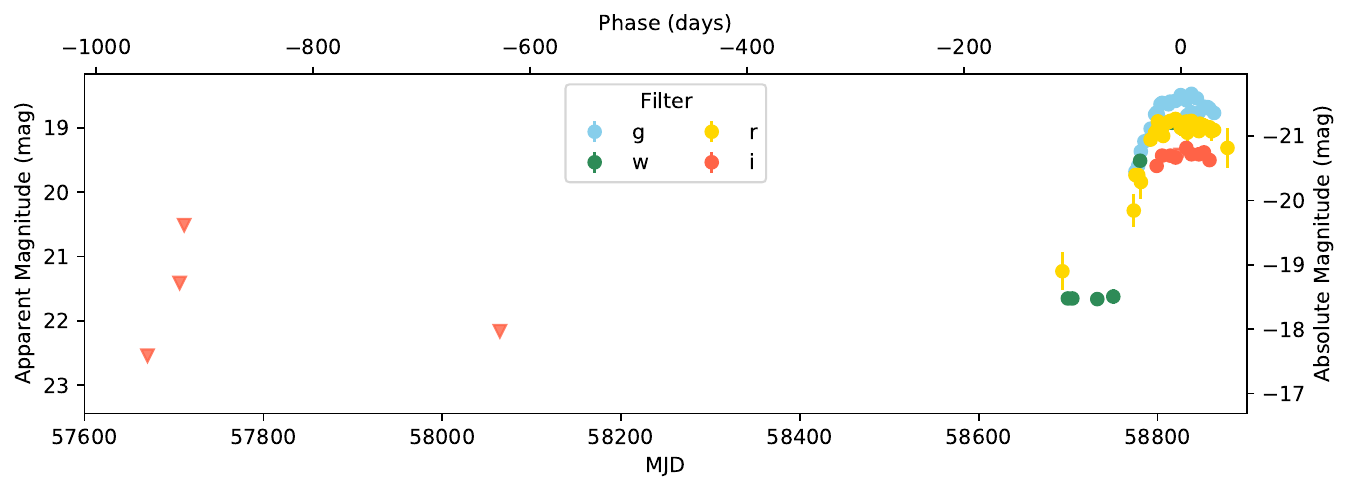}
    \caption{The early light curve of SN2019szu in $i$ and $w$ from Pan-STARRS forced photometry, and $g$ and $r$ from ZTF forced photometry. All magnitudes are in given in the AB system and 3$\sigma$ upper limits are indicated via inverted triangles. Phase given in rest frame days with respect to maximum light in the $g$-band. 
    }
    \label{fig:full iw lc}
\end{figure*}

\section{Observations}
\label{sec:observations}
\subsection{Discovery and Classification}

The Asteroid Terrestrial-impact Last Alert System (ATLAS) project \citep{Tonry2018} discovered SN2019szu with the designation ATLAS19ynd on 2019-10-21. The ATLAS-HKO (Haleakala) unit detected the supernova in the $c$ band at 19.4 mag following a shallow non-detection 2 days prior at a limiting magnitude of $o=17.7$ mag \citep{Tonry2019}. The transient was identified on multiple images and as it was coincident with a faint host galaxy (see Section\,\ref{sec:host}), the ATLAS Transient server reported it as a SN candidate \citep{Smith2020}. An earlier detection was made by the Zwicky Transient Facility (ZTF) on 2019-10-19 under the name ZTF19acfwynw \citep{Bellm2019}, with the data visible in the Lasair broker
\footnote{https://lasair-ztf.lsst.ac.uk/object/ZTF19acfwynw/} \citep{Smith2019}.
Gaia also detected this transient on 2019-11-02 with an internal name Gaia19fcb \citep{Wyrzykowski2016}. It was later classified as a SLSN-I by \citet{Nicholl2019a} as part of the C-SNAILS survey at the Liverpool Telescope (LT). It was initially given this classification using the host galaxy redshift of $z=0.213$ (based on using the [O III] doublet emission at 4959\,$\Angstrom$ and 5007\,$\Angstrom$ from the SLSN spectrum), and therefore an absolute magnitude $M=-21$ mag indicating a very luminous event. This redshift corresponds to a distance of $d=1060$ Mpc assuming a Planck cosmology \citep{PlanckCollaboration2020}. The absolute magnitude coupled with the very blue shape of the spectral continuum, the dwarf nature of the host galaxy and its strong emission lines similar to other SLSN hosts \citep{Leloudas2015b} cemented the SLSN designation. However, since the initial spectrum did not cover H$\alpha$, later spectra were needed to confirm its lack of hydrogen and type I designation. SN2019szu was also included as part of a large population study by \citet{Chen2023a, Chen2023b}. This sample consisted of 78 H-poor SLSNe detected by ZTF over the span of 3 years.

\subsection{Photometry}
\label{sec:photometry} 
Observations of this target were obtained from a number of telescopes. Follow-up observations in \textit{gri} were obtained with Las Cumbres Observatory (LCO) using a number of their 1m telescopes across multiple observatories in the network. After ~350 days, deeper images were obtained with the ESO New Technology Telescope (NTT) in \textit{gri} with the ESO Faint Object Spectrograph and Camera (EFOSC2) as part of the Extended Public ESO Spectroscopic Survey of Transient Objects \citep[ePESSTO;][]{Smartt2015}. 

These images were reduced using the ePESSTO pipeline for the NTT images \citep{Smartt2015}, and the BANZAI\footnote{https://github.com/LCOGT/banzai} pipeline for the LCO images. Photometry on these images was performed with the use of \textsc{photometry sans frustration}, a python wrapper for point-spread function (PSF) photometry using Astropy and Photutils \citep{Nicholl2023}. Zeropoints were calculated by cross matching sources in the field with the Pan-STARRS catalog \citep{Flewelling2020}. The photometry in $gri$ was template subtracted using archival PS1 images as templates. This was especially important in the late time photometry where we believe the host plays a significant contribution to the flux detected.

The Neil Gehrels Swift Observatory (hereafter \citealp[\textit{Swift};][]{Gehrels2004}) began observations of the field of SN2019szu on the 21st November 2019 using the UV Optical Telescope \citep[UVOT;][]{Roming2005}. SN2019szu was detected in all 6 optical/UV UVOT filters. Summed images were created by combining individual exposures taken during observations. Source counts were extracted from these summed images using a source region of 5" radius. Background counts were extracted using a circular region of radius 20" located in a source-free region. The count rates were obtained from the image lists using the {\it Swift} tool {\tt uvotsource}. The count rates were then converted to AB magnitudes using the UVOT photometric zero points \citep{Poole2008, Breeveld2011}.

Data were also collected from the ZTF forced photometry server \citep{Masci2019} and the ATLAS forced photometry server \citep{Tonry2018, Smith2020, Shingles2021}, both of which are performed on difference images. ZTF $g$, $r$, and ATLAS $c$ band data points were binned together in a daily cadence whereas the $o$ band data were binned together every 2 days to reduce noise.

Following the discovery of SN2019szu, we examined forced photometry at the location of the SN in Pan-STARRS1 and Pan-STARRS2 images obtained in survey mode \citep{2016arXiv161205560C} from MJD 57362 (2015-12-06) onwards. Typically,  $4\times45$ second exposures are obtained in survey mode in one of $w$, $i$ or $z$ filters on any given night, and photometric calibration and difference imaging is performed via the image processing pipeline \citep[IPP;][]{2020ApJS..251....3M}. The $w$ filter is a broadband composite $g+r+i$, with measured AB magnitudes roughly equivalent to those in the $r$ band. The individual measurements for each nightly quad were stacked in order to improve signal to noise, and to obtain deeper upper limits in case of a non-detection.

All absolute magnitudes are calculated using the distance modulus and a simple K-correction of 2.5\,log($1+z$).

\subsection{Polarimetry}

An epoch of polarimetry was also obtained on 17-01-2020 (31 days after peak in rest frame) using the Alhambra Faint Object Spectrograph and Camera (ALFOSC) instrument at the Nordic Optical Telescope (NOT) in the V band. The reduction and analysis is described in \citet{Pursiainen2023}. The signal-to-noise ratio (S/N) is quite low compared to values traditionally needed for linear polarimetry. We find S/N $\gtrsim 100$ only for a small aperture size of $\le9$ pixels, and falls quickly to ~35 for aperture sizes above 20 pixels (larger apertures are necessary to account for any difference in point spread function between the ordinary and extraordinary beams). Although we measure an overall polarisation of $P=3.1\pm1.3\%$ for SN2019szu, compared to a an interstellar polarisation $P_{\rm ISP} = 0.70\pm0.21\%$ measured from a bright nearby star, the low S/N of the observation precludes a strong claim of polarized emission from SN2019szu. This is described in detail by \citep{Pursiainen2023}.

\begin{table*}
	\centering
	\caption{Spectroscopic observations of SN2019szu. Phase is given in rest frame frame days with respect to the time of maximum light in the $g$-band. $^{*}$Identical arms.}
	\label{tab:spectra}
	\begin{tabular}{cccccccc} 
		\hline \hline
		Date & MJD & Phase & Telescope & Instrument & Grism/Grating & Exposure time (s) & Wavelength Range ($\Angstrom$)\\
		\hline
		02-11-2019 & 58789 & -30 & LT & SPRAT & Blue grating & 1800 & 4000-8000 \\
		20-11-2019 & 58807 & -16 & NTT & EFOSC2 & Gr\#13 & 1800 & 3685-9315 \\
            27-11-2019 & 58814 & -10 & NTT & EFOSC2 & Gr\#13 & 2700 & 3685-9315 \\
            07-12-2019 & 58824 & -2 & NTT & EFOSC2 & Gr\#13 & 2700 & 3685-9315 \\
            19-12-2019 & 58836 & 7 & NTT & EFOSC2 & Gr\#13 & 2700 & 3685-9315 \\
            02-01-2020 & 58850 & 18 & NTT & EFOSC2 & Gr\#13 & 2700 & 3685-9315 \\
            16-01-2020 & 58864 & 30 & NTT & EFOSC2 & Gr\#13 & 735 & 3685-9315 \\
            17-01-2020 & 58865 & 31 & NTT & EFOSC2 & Gr\#16 & 1500 & 6015-10320  \\
            21-08-2020 & 59082 & 211 & MMT & Binospec & 270 lpmm$^{*}$ & $5\times800$ &  3900-9240 \\
            23-10-2020 & 59145 & 262 & NTT & EFOSC2 & Gr\#13 & 2700 & 3685-9315 \\
		\hline \hline
	\end{tabular}
\end{table*}

\subsection{Spectroscopy}
\label{sec:spectroscopy} 
An initial spectrum of SN2019szu was obtained using the Spectrograph for the Rapid Acquisition of Transients (SPRAT) instrument on the Liverpool Telescope (LT). Spectroscopic follow-up observations of this target were then undertaken by ePESSTO using the NTT with EFOSC2 \citep{Smartt2015}. Most of these spectra were obtained with Gr$\#$13 except for one epoch with Gr$\#$16 on 2020-01-17 to extend the wavelength coverage redwards. The latter was averaged in the overlapping region with the Gr$\#$13 spectrum obtained one day prior. Another spectrum was obtained on 2020-08-21 from MMT using the Binospec spectrograph covering a similar wavelength range to Gr$\#$13 \citep{Fabricant2019}. A full breakdown of observations is given in Table \ref{tab:spectra}.

All data were reduced using dedicated instrument-specific pipelines that apply de-biasing, flat-fielding, trace extraction, wavelength calibration and flux calibration using standard stars observed with the same setup. Spectra were then also flux corrected using both $r$ and $i$-band photometry.

\section{Analysis}
\label{sec:analysis}

\subsection{Host Galaxy}
\label{sec:host} 
The host galaxy of SN2019szu is a faint dwarf detected in the Pan-STARRS catalogue (PSO J002.5548-19.6923). There is no catalogued redshift or distance information and so estimates for the redshift were derived from the [OIII] $\lambda\lambda$4959,5007 narrow host lines observed in the SN spectra. This gave a redshift of $z=0.213\pm0.0003$ which is also in agreement with H$\alpha$ and H$\beta$ measurements from latter spectra.

The host is detected only in the Pan-STARRS $r$ and $i$ bands with a Kron magnitude $r=21.75 \pm 0.08$\,mag \citep{Flewelling2020}. It is not catalogued in the NASA Extragalactic Database \citep{Helou1991}. In \citet{Schulze2018}, the median SLSN host galaxy had $M_{B} = -17.10 \pm 1.45$ mag. The host for SN2019szu has an absolute magnitude $M_{r} = -18.38 \pm 0.08$ mag (at this redshift, the $r$-band is similar to rest-frame $B$). This indicates the SN2019szu host galazy is well within the normal range of host luminosities (a proxy for masses) found in \citet{Schulze2018}.

Hydrogen line ratios were measured based on the narrow emission lines observed in the late-time spectra of SN2019szu in order to estimate any reddening due to the host galaxy. This gave a ratio of H$\alpha/$H$\beta = 3.53 \pm 0.13$  for the spectrum at +211 days, and H$\alpha/$H$\beta = 4.52 \pm 0.45$ for the spectrum at +262 days. The H$\gamma$/H$\beta$ value could only be calculated for the +211 day spectrum and resulted in a value of H$\gamma/$H$\beta = 0.73 \pm 0.17$. Both of these values are above the expected ratios of H$\alpha/$H$\beta = 2.86$, and H$\gamma/$H$\beta = 0.47$ \citep{Osterbrock2006}. While the H$\gamma/$H$\beta$ ratio supports negligible extinction, the H$\alpha/$H$\beta$ ratio indicates significant reddening from the host, which is unexpected for a galaxy of this size. We can quantify the relation between Balmer decrement and colour excess $E(B-V)$ described by \citet{Dominguez2013} giving $E(B-V) = 0.29 \pm 0.05$ and optical extinction of $A_{V} = 1.2 \pm 0.3$. This is much larger than typically inferred for SLSN host galaxies, which is generally $<$0.5 mag and averages $\sim$0.1 mag \citep{Schulze2018}. It is therefore likely that our measured Balmer decrement is unreliable, due to contamination from the SN spectrum. A low host extinction is also supported by light curve models (Section \ref{sec:magentar models}) and the lack of NaI D $\lambda\lambda$5890, 5896 absorption which is used as a indicator of dust extinction \citep{Poznanski2012}. We therefore neglect host extinction in our analysis, as applying more host extinction didn't affect our spectral measurements significantly. The Milky Way extinction in the direction of SN2019szu is $E(B-V)=0.018$ \citep{Schlafly2011}, and this correction was applied to all spectra.

The metallicity of the galaxy was calculated from the $R_{23}$ method outlined in \citet{Kobulnicky1999}, which uses the fluxes of the  [O II] $\lambda$3727, [O III] $\lambda\lambda$4959,5007, and H$\beta$ lines. As the metallicity appeared to be in the region between the metal rich and the metal poor branches of this relation, it was calculated for both branches. The metal rich branch yields a value of $12+\text{log(O/H)}=8.36\pm0.07$, and the metal poor branch a value of $12+\text{log(O/H)}=8.36\pm0.05$. Both of these values are in agreement with one another, and also consistent with work by \citet{Schulze2018} who found SLSNe host galaxies tended to have values of  $12 + \text{log(O/H)} < 8.4$. Limits on the metallicity can also be placed by measuring the ratio of [N II] $\lambda$6583 to H$\alpha$. [N II] is not observable in SN2019szu and so using a 2$\sigma$ and 3$\sigma$ limit on this detection yields $12 + \text{log(O/H)} < 8.06$ and $12 + \text{log(O/H)} < 8.14$ respectively \citep{Marino2013}. This reinforces the low metallicity nature of the host.

The host for SN2019szu shows strong H$\alpha$ and [O III] $\lambda$5007 emission lines. In the +211 day Binospec sepctrum, these lines have equivalent widths of 148\,$\Angstrom$\ and 127\,$\Angstrom$\ respectively which are lower limits due to contamination from the SN continuum. This is similar to the sample of SLSN host galaxies studied by \citet{Leloudas2015b}, where $\sim$50\% of the sample occurred in extreme emission line galaxies.

Using the star formation rate (SFR) diagnostics in \citet{Kennicutt1998} provides two different measures of the SFR. The first uses the strength of  H$\alpha$ and gives a value of SFR $= 0.4 $\,M$_{\odot}$ yr$^{-1}$. The second uses [O II] $\lambda$3727 and gives SFR $= 0.3-0.4 $\,M$_{\odot}$ yr$^{-1}$. Comparing to the sample of SFRs found in \citet{Leloudas2015b} which ranged from $0.01-6.04$\,M$_{\odot}$ yr$^{-1}$, this is a typical star formation rate for SLSNe hosts.

\subsection{Light Curve}
\label{sec:light curve} 

The multi-band light curve of SN2019szu is shown in Figure \ref{fig:light curve split} with a range spanning over 600 days in the observer frame. The event reaches a peak magnitude of $M_{g,\text{peak}}= -21.59\pm0.06$, which is close to the volume corrected median peak magnitude $M_{\text{peak}} = -21.31\pm0.73$ mag described by \citet{Lunnan2018b}, and $M_{g,\text{peak}} = -21.14\pm0.75$ mag calculated by \citet{DeCia2018} for SLSNe, where the error represents the 1$\sigma$ spread. A more recent study by \citet{Chen2023a} found $M_{g,\text{peak}} = -21.54^{+1.12}_{-0.61}$ mag (not corrected for Malmquist bias).

The main rising light curve is preceded by a plateau lasting 40 days in the rest frame. This commences at MJD 58700, and hovers around $w=21.6$\,mag before beginning to rise after MJD 58750. This is equivalent to an absolute magnitude of M$_{w}\sim-18.7$\,mag and a luminosity $\sim10^{43}$\,erg\,s$^{-1}$. This plateau was also observed in the $r$-band. Figure \ref{fig:full iw lc} shows historical photometry in the $i$-band showing deep upper limits down to 22.5\,mag indicating this plateau is a emergent feature. Other SLSNe have shown signs of early excesses such as SN2006oz, in which a precursor plateau lasting 10 days was observed before the full monotonic rise. This was thought to be due to a recombination wave in the surrounding CSM consistent with the transition from O III to O II \citep{Leloudas2012}. Similarly SN2018bsz showed a slowly rising plateau lasting $\sim30$ days \citep{Anderson2018}. In LSQ14bdq the precursor peak was suggested to be caused by the cooling of extended stellar material \citep{Nicholl2015b}. In both of these cases the precursor events may have occurred after explosion, unlike in SN2019szu where the long plateau clearly precedes the explosion date inferred from the rising light curve. This feature in SN2019szu is also unusual due to its very flat nature over a long timescale which is not consistent with cooling material and may require an additional source of energy injection.

This light curve rise is captured well by ZTF and ATLAS. We estimated the date of maximum light to be MJD 58826 in the $g$ band by fitting a low-order polynomial, giving a rise time of around 80 days. We take this to be the time of peak throughout. We caution that our fits to the peak are somewhat limited by SN2019szu entering solar conjunction around 50 days after discovery. Observations resumed when it became visible again around 90 days later. 

The light curve appears to peak in the bluer bands first and has a rather flat shape or possible plateau at peak, which lasts longer in redder bands. In the $r$ band this flattening lasts approximately 80 days. This is similar to other events such as SN2020wnt which also showed this plateau behaviour \citep{Gutierrez2022, Tinyanont2023}. However, SN2020wnt also showed indications of an initially faster decline in bluer bands, which is not apparent in this event (Figure \ref{fig:light curve split}). The UV bands all show rising light curves until the solar conjunction, indicating a peak later in the evolution for these bands.

To parameterise the light curve peak, the exponential rise and decline timescales were determined giving an e-folding rise and decline of $\tau_{g-\text{rise}} \sim$ 48 days, and $\tau_{g-\text{decline}} \sim$ 100 days respectively. These timescales were determined by fitting low order polynomials to the g band light curve. In general SLSNe tend to have $\tau_{\text{rise}} \sim \tau_{\text{decline}} / 2$ with slower evolving events also having slower rise times \citep{Nicholl2015a}. The rise and decline timescales of SN2019szu are consistent with this expectation. Some events however show a more skewed relation such as SN2017egm, which had a fast rise time $\tau_{g-\text{rise}} \sim$ 20 days, and slow decline with an estimated e-folding decline time of $\tau_{g-\text{decline}} \sim$ 60 days \citep{Bose2018}.

A small bump in the light curve can be seen around MJD 59100, corresponding to $\sim$200 days post peak. This is not unusual for SLSNe and a large fraction of SLSNe-I show these undulations \citep{Hosseinzadeh2022,Chen2023b}. This undulation is observed in the \textit{gr} bands, which have sufficient coverage to observe variations at this phase. As discussed in \citet{Nicholl2016b, Inserra2017, Li2020}, these undulations can be the result of collisions of the ejecta with shells or clumps of material. This interaction with circumstellar material can produce bluer colours ($g-r$) during the interaction due to heating of the ejecta \citep{Chen2023b}. An alternative theory is variation in the power output from a central engine such as the energy output from a magnetar \citep{Metzger2018, Chen2023b}. Although this would produce variability on timescales shorter than the observed bumps, this can be smoothed at early times if the variation is shorter than the photon diffusion time through the ejecta. This also implies that the undulations are more likely visible at later times as the ejecta becomes more transparent -- this is supported by the fact that 73$\%$ of undulations are found post peak \citep{Chen2023b}. A central engine can also produce variations in the ejecta opacity via increased ionisation and hence more electron scattering even with constant energy input. It does this by creating an ionisation front that propagates outwards and breaks out from the front of the ejecta leading to a rebrightening \citep{Metzger2014, Omand2023}. As the ejecta cools it can recombine, leading to a change in opacity again which could result in an observed undulation. \citet{Chen2023b} prefer this latter explanation for SN2019szu as this mechanism allows for a higher UV flux due to the decrease in bound-bound transitions for ionised metals. 

Late time observations of SN2019szu (Figure \ref{fig:full iw lc} show deep upper limits in both $i$ and $w$, at levels below the precursor plateau observed. This supports the idea that this plateau is related to the SN event.

\begin{figure}
	\includegraphics[width=\columnwidth]{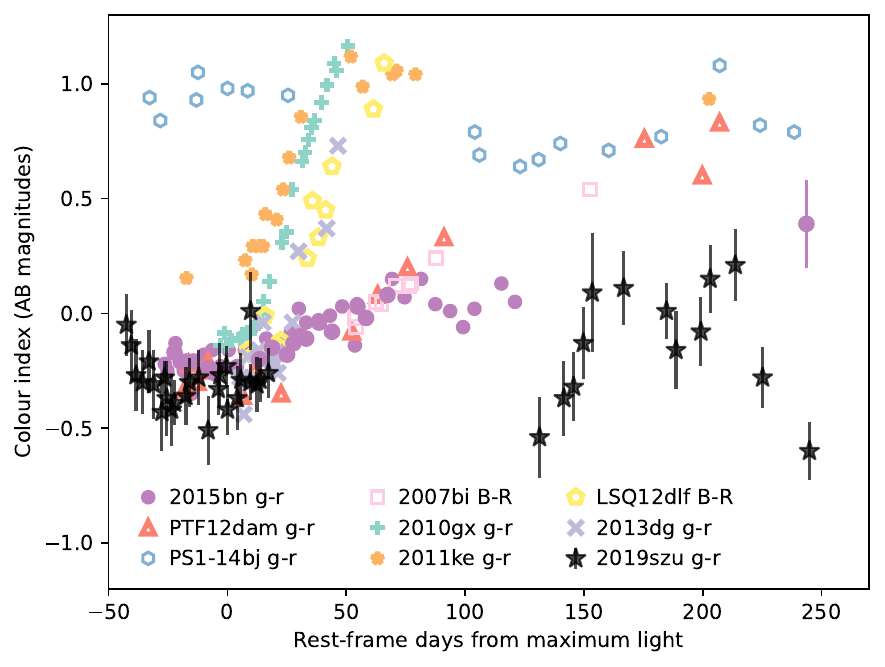}
    \caption{Colour evolution of SN2019szu in $g-r$ compared a subset of SLSNe including SN2007bi, PTF12dam, and SN2015bn which all showed early emission of the 7300\,$\Angstrom$ line in their spectra \citep{Gal-Yam2009, Nicholl2013, Nicholl2016b}. PS1-14bj is also included due to its extremely slow evolution \citep{Lunnan2016}. Other well observed SLSNe with published \textsc{Superbol} fits and more typical colour evolutions have also been plotted including SN2010gx, SN2011ke, LSQ12dlf, and SN2013dg \citep{Pastorello2010, Inserra2013b, Nicholl2014}. All photometry has been K-corrected.}
    \label{fig:colour}
\end{figure}

\subsubsection{Colour}
\label{sec:colour}

The colour evolution of SN2019szu is shown in Figure \ref{fig:colour}. The colour was calculated using \textsc{Superbol}, a python package that interpolates light curves in order to perform spectral energy distribution (SED) fits \citep{Nicholl2018a}. The $g-r$ colour was calculated using our $g$-band data, and interpolated $r$-band points from \textsc{Superbol} using a polynomial fit. The pre-peak colour shows a dramatic evolution to the blue, from an initial value of $g-r=0.05\pm0.13$ mag, dropping down to $g-r=-0.51\pm0.15$ mag just before maximum light. This is not behaviour exhibited by other SLSNe, which tend to show a general colour increase, becoming redder over time shown by the sample of events in Figure \ref{fig:colour}. The colour of SN2019szu, exhibits only a very gradual change in colour after peak, hovering around $g-r\sim-0.3$ mag. This is similar to other SLSNe which show a steady blue colour around peak before evolving dramatically towards the red, consistent with fast cooling after peak. Slow SLSNe such as SN2007bi, PTF12dam, and SN2015bn show a much more gradual colour evolution to the red \citep{Gal-Yam2009, Nicholl2013, Nicholl2016b}, similar to the post peak evolution of SN2019szu. PS1-14bj is another event showing near constant colour but is overall a much redder event \citep{Lunnan2016}.    

After the break in data, the colour appears to redden, consistent with slow cooling as seen in other SLSNe. At around 200 days post peak the colour once again begins to decrease. This turning point corresponds to the bump in the light curve mentioned in Section \ref{sec:light curve}. This observation could be explained by an ionisation front breaking out of the ejecta \citep{Metzger2014, Omand2023}, or interaction with circumstellar material \citep{Inserra2017}. Both of these mechanisms would heat the ejecta and therefore create a bluer colour. This is not something seen in other bumpy light curves, for example SN2015bn does not have a dramatic colour change around its bump at +50 days \citep{Nicholl2016b}. We also caution that the late time $g-r$ colour could be affected by the strong emission lines from the host found in the $r$-band wavelength range.

\subsubsection{Bolometric Luminosity}
\label{sec:bol lum}

\begin{figure}
	\includegraphics[width=\columnwidth]{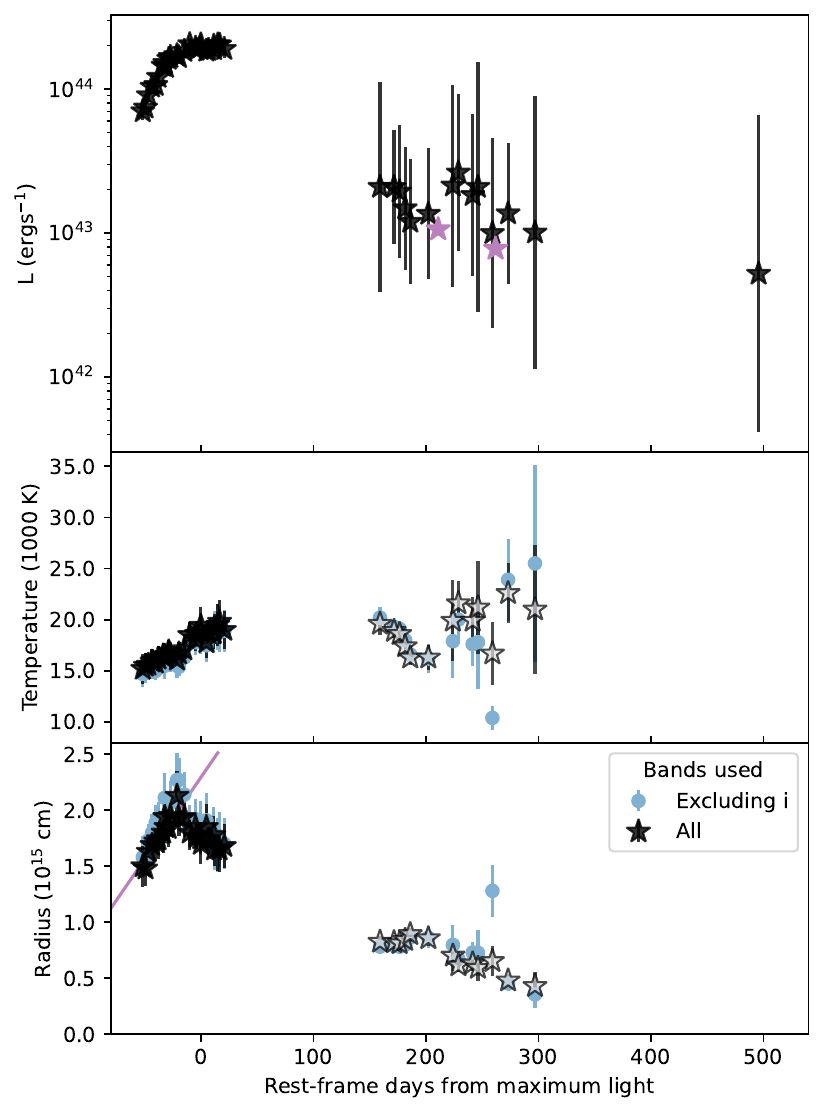}
    \caption{Measurements from fitting blackbodies to SEDs constructed from \textsc{Superbol}. \textbf{Top:} Bolometric luminosity of SN2019szu constructed using \textit{uvw1, uvm2, uvw2}, and \textit{Ugcwroi} up until the initial break (\textit{gcroi} thereafter). Purple stars represent the luminosity derived by integrating under the +211 and +262 day spectra. 
    \textbf{Middle:} Blackbody temperature of SN2019szu calculated with black stars using all bands and blue points excluding $i$.
    \textbf{Bottom:} Blackbody radius of SN2019szu calculated with black stars using all bands and blue points excluding $i$. The radius is fit (purple line) to the initial rise up to -20 days relative to peak and indicates and expanding photosphere at $\sim$1200\,\kms. After solar conjunction, points are unfilled to represent sparse photometry and large uncertainties.}
    \label{fig:bb fits}
\end{figure}

The bolometric luminosity of SN2019szu was calculated using \textsc{Superbol} \citep{Nicholl2018a}, as shown in Figure \ref{fig:bb fits}. To do this the light curve in each band was interpolated to epochs with $g$-band data. A constant colour relation was assumed for bands with fewer data points, or by fitting a low-order polynomial to capture the general shape of the light curve. The flux was corrected for time-dilation assuming a redshift of $z=0.213$, and extinction corrected assuming a value of $E(B-V)=0.018$ \citep{Schlafly2011}. As discussed in Section \ref{sec:host}, we assume negligible extinction from the host galaxy. The data was also K-corrected to shift the fluxes and effective filter wavelengths to their rest-frame values. The resulting spectral energy distribution (SED) was fit with a modified blackbody function suppressed below a cut-off wavelength \citep{Nicholl2017a, Yan2018}:
\begin{equation}
    f_{\lambda}(T, R) = 
    \begin{cases}
      (\frac{\lambda}{\lambda_{0}})^{\beta} f_{\lambda, \text{BB}}(T, R) \text{\qquad \qquad \qquad for $\lambda < \lambda_{0}$} \\
      f_{\lambda, \text{BB}}(T, R) \text{\qquad \qquad \qquad \qquad \enspace for $\lambda > \lambda_{0}$}
    \end{cases}
    \label{eq:SED}
\end{equation}
Where $f_{\lambda}$ is the wavelength dependent flux, $\lambda$ is the wavelength, $\beta$ is a nominal index for which we used a value of 3, and a cutoff wavelength of $\lambda_{0}$=3000\,\AA\ was used. These values were chosen based on fitting Eq. \ref{eq:SED} to each SED and averaging the best fits. Multicolour information was not available for the pre-explosion plateau and so this data was not included in the SED fitting. 

\begin{figure*}
	\includegraphics[width=2\columnwidth]{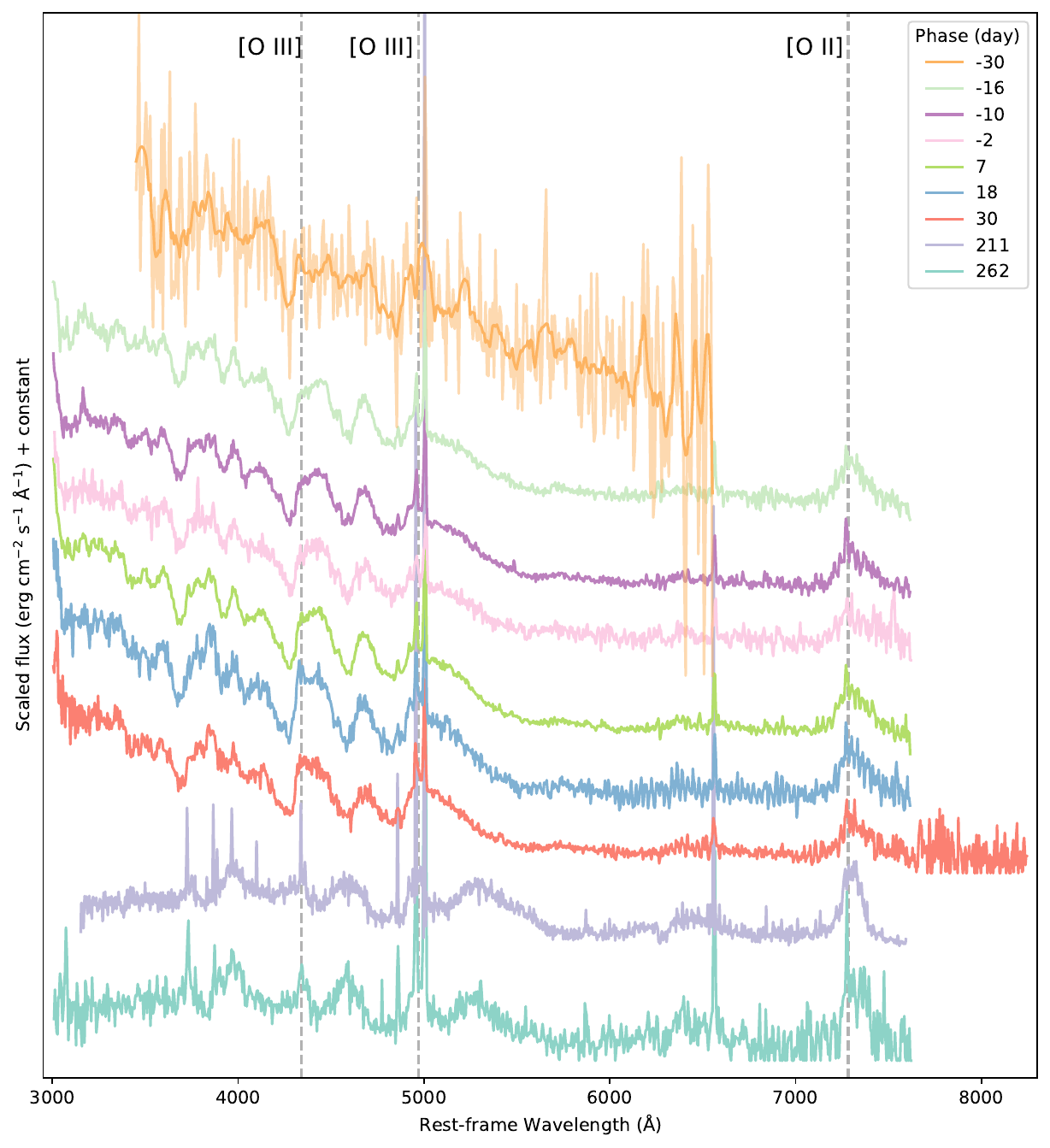}
    \caption{Spectral evolution of SN2019szu, phase is given in rest frame days with respect to maximum light in the $g$-band. Dashed vertical lines indicate blue shifted positions of ionised oxygen lines seen in the spectra. The first spectrum at $-$30 is presented as both a smoothed and unsmoothed version for clarity. Edges of of this spectrum are clipped to remove noisy edges. Spectrum at 30 days is a composite of spectra taken on 30 and 31 days.  The spectrum at 211 days has been telluric corrected and also smoothed using a Savitzky–Golay filter to reduce noise.}
    \label{fig:spec ev}
\end{figure*}

Fitting this equation using data from all bands produced the blackbody (BB) temperature (\textit{T}) and radius (\textit{R}). \textsc{Superbol} calculates the bolometric luminosity ($L_{\rm{bol}}$) by integrating numerically under the observed SED points, and extrapolating the missing flux outside the wavelength range using the best-fitting absorbed BB model (Figure \ref{fig:bb fits}). The initial peak is fit using data from the \textit{uvw1, uvm2, uvw2}, and \textit{Ugcwroi} filters; the $B$ and $V$ bands were ignored due to their very sparse data points. Both bands were also very noisy and so did not provide much extra information compared to the much cleaner $g$ band. After the first break due to solar conjunction, only \textit{gri} data was used in order not to extrapolate too far in time in the other bands, instead opting to extrapolate further in wavelength. As we will discuss in section Section \ref{sec:continuum}, the SED shape appears flatter than a blackbody in the redder bands. We therefore performed additional fits excluding the \textit{i} band. These experiments showed no significant difference to the best-fit $L_{\rm{bol}}$, but did marginally affect $T$ and $R$ (Figure \ref{fig:bb fits}). We also caution that a late times (>200 days) the SED does not resemble a blackbody as evidenced by the nebular spectra in Figure \ref{fig:spec ev}, and so the bolometric luminosity should be treated with a degree of caution.

$T$, $R$ and $L_{\rm{bol}}$ were calculated for each epoch, however, it is important to note that at later times the blackbody fits are more contaminated by nebular lines and as the event transitions from photopheric to nebular, the blackbody fit becomes less reliable. Beyond 300 days, we have detections in only one band, which is insufficient to measure a temperature and radius. However, assuming that the colour does not change dramatically, we are able to estimate the bolometric luminosity at the time of this last detection.

\begin{figure}
	\includegraphics[width=\columnwidth]{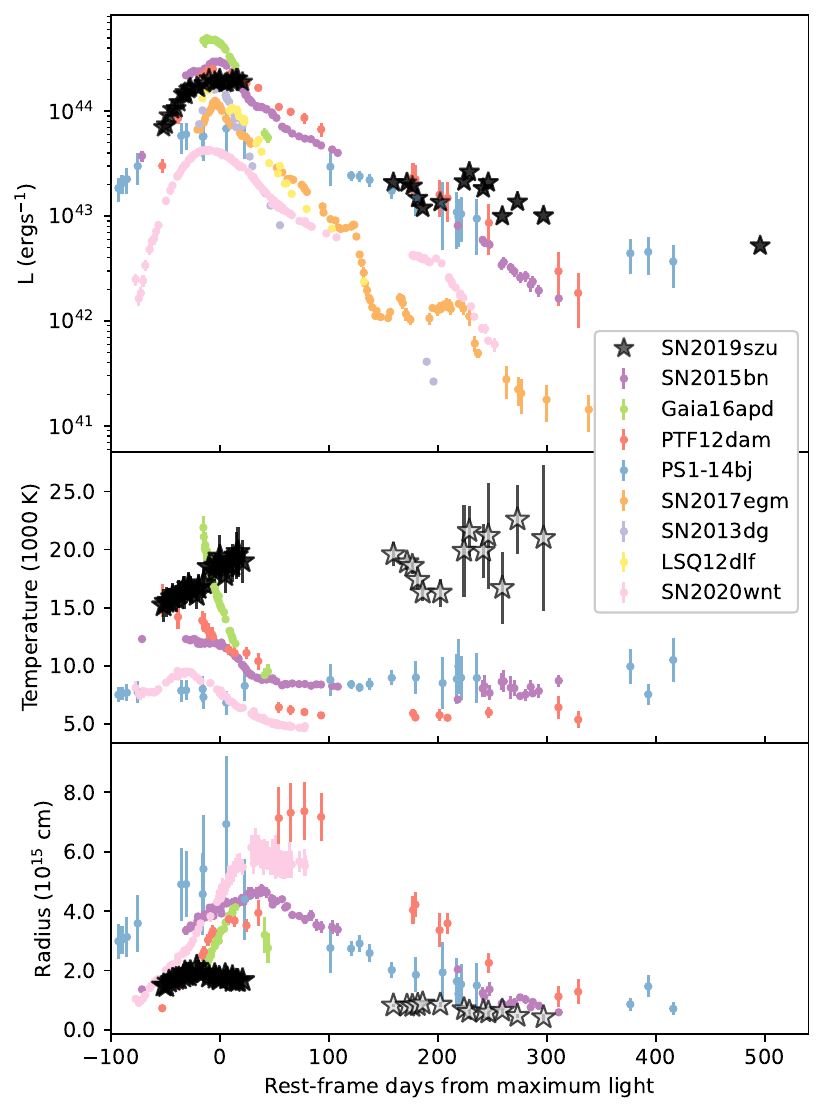}
    \caption{\textbf{Top:} Bolometric luminosity of SN2019szu compared to a subset of well observed SLSNe with published \textsc{Superbol} fits. Error bars of SN2019szu have been removed for clarity. \textbf{Middle:} Blackbody temperature of SN2019szu calculated using \textsc{Superbol}, compared to other SLSNe. \textbf{Bottom:} Blackbody radius of SN2019szu calculated using \textsc{Superbol}, compared to other SLSNe. After solar conjunction, points are unfilled to represent sparse photometry and large uncertainties. \citep{Nicholl2016b,Nicholl2016c,Nicholl2013,Lunnan2016,Lin2023,Bose2018,Nicholl2017c,Nicholl2014,Gutierrez2022}}
    \label{fig:bb comparisons}
\end{figure}

In Figure \ref{fig:bb comparisons}, we can see our event compared to some other well observed SLSNe with published \textsc{Superbol} data. Even compared to these other events, the slow nature of this event is apparent with only PS1-14bj having a comparable gradual decline. It is much harder to compare the rise to peak of these events as they are not all as well sampled, but we note that SN2019szu has a much faster rise to peak than PS1-14bj.

Integrating the bolometric light curve gives us an energy of $E=2.6\times10^{51}$\,erg. This is a lower limit on the energy radiated by this event due to our finite sampling in time and wavelength. It also does not consider the energy released outside of time span covered by our photometry. The energy is consistent with other SLSNe which typically radiate $\sim10^{51}$\,erg over their lifetimes \citep{Gal-Yam2012,Lunnan2018b}.

\begin{figure*}
	\includegraphics[width=2\columnwidth]{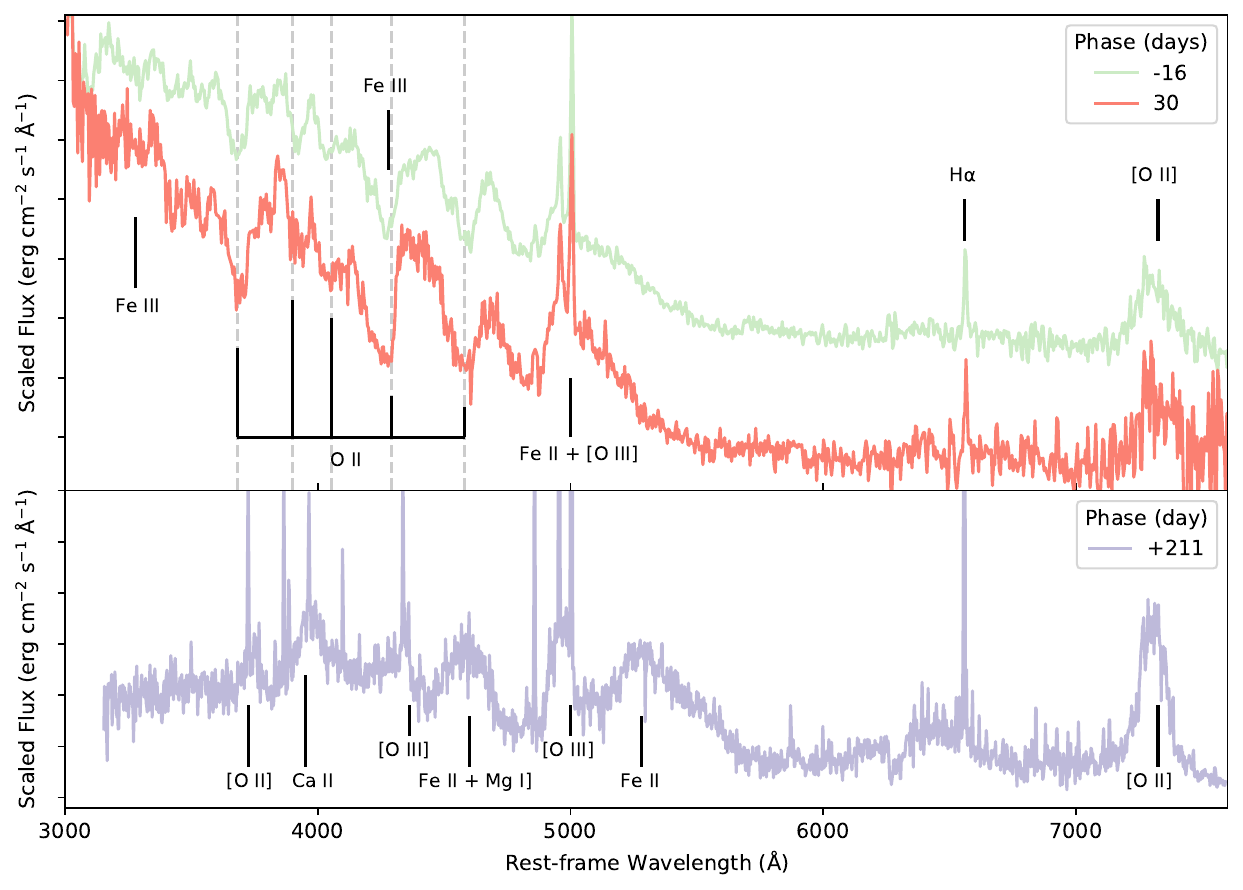}
    \caption{Spectra of SN2019szu with the main features labelled. Phase given in rest frame days with respect to maximum light in the $g$-band. \textbf{Top:} Photospheric spectra with predominantly SLSNe absorption lines and a visible emission line attributed to [O II] at $\lambda$7320. The narrow components of the [O III] lines and the narrow H$\alpha$ are attributed to the host-galaxy. Dashed vertical lines indicate the O II absorption lines \citep{Quimby2018}, if blueshifted with a velocity of 4500\,\kms. \textbf{Bottom:} Nebular spectra with SLSN emission lines. Narrow lines are attributed to the host-galaxy. Y axis cut for clarity.}
    \label{fig:labelled}
\end{figure*}

\subsubsection{Blackbody Temperature and Radius}
\label{sec:BBrandt}

Figure \ref{fig:bb comparisons} shows the blackbody temperature and radius compared to other SLSNe, calculated using the method discussed in \ref{sec:bol lum}. Although this method allowed us to extrapolate the SED outside of the observed bands and measure the total luminosity, the continuum shape visible in our spectra clearly deviates from a blackbody even at early times as seen in Figure \ref{fig:spec ev}. This will be discussed in Section \ref{sec:continuum}, but for our purposes here this means that forcing a blackbody fit on this data may result in temperature and radius measurements that are not physically meaningful. As the deviations become apparent at rest-frame wavelengths longer than 5500\,$\Angstrom$, these SED fits were repeated with the \textit{i} band removed. Removing this band did not produce any changes in temperature or radius larger than the 1$\sigma$ uncertainties on these quantities at early times, however some late time points between 200-300 days show significant deviations. The evolution excluding the \textit{i} band shows significant fluctuations in short timescales for both temperature and radius, as well as much larger uncertainties. For this reason the fits including the $i$ band were chosen for further analysis.

By looking at Figure \ref{fig:bb comparisons}, it is apparent that the temperature evolution is not consistent with other SLSNe, as it appears to increase over the first $\sim$100 days. This is in stark contrast with most other events, which tend to show a decreasing temperature \citep{Chen2023a}. However, it is consistent with the colour evolution found in Section \ref{sec:colour} which indicated the SN became bluer with time around peak. At later times the temperature of SN2019szu has larger errors which makes it harder to constrain but it appears as though the SLSN stays much hotter than the other events, and also remains roughly constant rather than increasing or decreasing drastically. SN2020wnt and PS1-14bj both show increases in temperature post explosion \citep{Gutierrez2022, Lunnan2016}. In SN2020wnt this increase is only before the peak of the light curve and lasts $\sim$50 days post explosion before decreasing in temperature. PS1-14bj instead shows a steady increase over the entire time frame, however both events stay cooler than SN2019szu. \citet{Lunnan2016} suggest that late-time heating, due to X-ray to UV breakout from a central engine, could explain this increase. We also note that the large discrepancy in the temperature evolution of SN2019szu compared to other events might suggest that indeed a blackbody is not an accurate representation of its SED.

\citet{Chen2023a} also analysed the ZTF light curve of SN2019szu in a population paper of 78 SLSNe-I observed from March 17, 2018 to October 31, 2020. That paper highlighted the anomalous nature of SN2019szu (designated ZTF19acfwynw). They presented the same increasing temperature profile that we found in Figure \ref{fig:bb fits}. \citet{Chen2023a} provide possible explanations of CSM interaction providing an additional heating source, or ejecta being ionised by a central engine such as a magnetar. Alternatively the apparently rising temperature may be due to mismatch between the true SED and the assumption of a thermal spectrum. These possibilities will be discussed in Section \ref{sec:discussion}. 

Looking at Figure \ref{fig:bb fits}, we can see the radius appears to peak at $R = (2.13 \pm 0.22) \times 10^{15}$ cm at around 20 days before maximum light. Afterwards, the radius appears to either decrease very slowly or remain relatively constant up to the break in the data. Fitting this initial rise up to -20 days with a linear function appears to show the photosphere expanding at $v \sim 1700$\,\kms. After the break the radius appears to have decreased significantly, where it continues a slow, gradual decline down to $R = (4.31 \pm 1.26) \times 10^{14}$ cm by 300 days. Overall the blackbody radius of SN2019szu appears quite compact compared to other SLSNe in our comparison sample, by a factor of few. Slower moving ejecta for SN2019szu could be one possible explanation for this difference. However, it may also be that the deviations of the spectrum from a true blackbody, as indicated by the apparently increasing temperature (and discussed further in the next section), lead to an underestimate of the true radius. This is supported by SN2018ibb which had an ejecta velocity of 8500\,\kms but displayed a steady photophere radius of $\sim5\times10^{15}$\,cm over the course of 100 days \citep{Schulze2023}, comparable to the radius of other events.

\begin{figure}
	\includegraphics[width=\columnwidth]{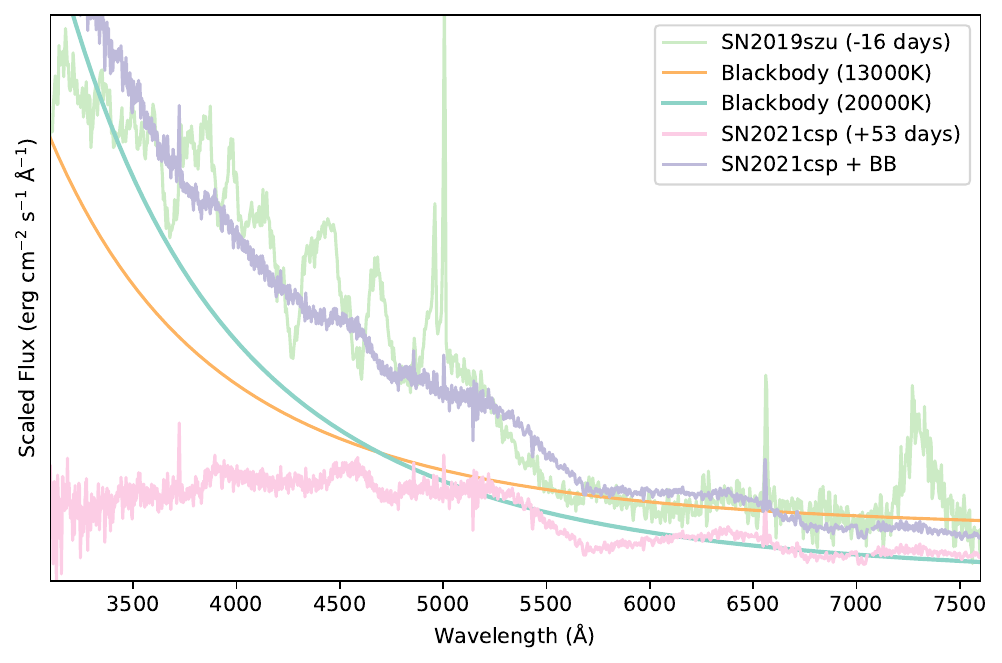}
    \caption{Comparisons between the spectrum of SN2019szu at -16 days as well a smoothed spectrum of interacting Type Icn SN021csp at +53 days, with a strong pseudo-continuum} \citep{Perley2022}. In orange is the best fit blackbody to SN2019szu with a temperature component $T \sim$ 13000\,K, shifted vertically to align with the flat red continuum. In blue is a scaled blackbody with a temperature component $T \sim$ 20000\,K. A composite spectrum is also shown created by summing the 20000\,K blackbody and SN2021csp spectrum (representing the interaction component) which can recreate both the flat red continuum and the steep blue continuum shape.
    \label{fig:best bb fit}
\end{figure}

\begin{figure}
	\includegraphics[width=\columnwidth]{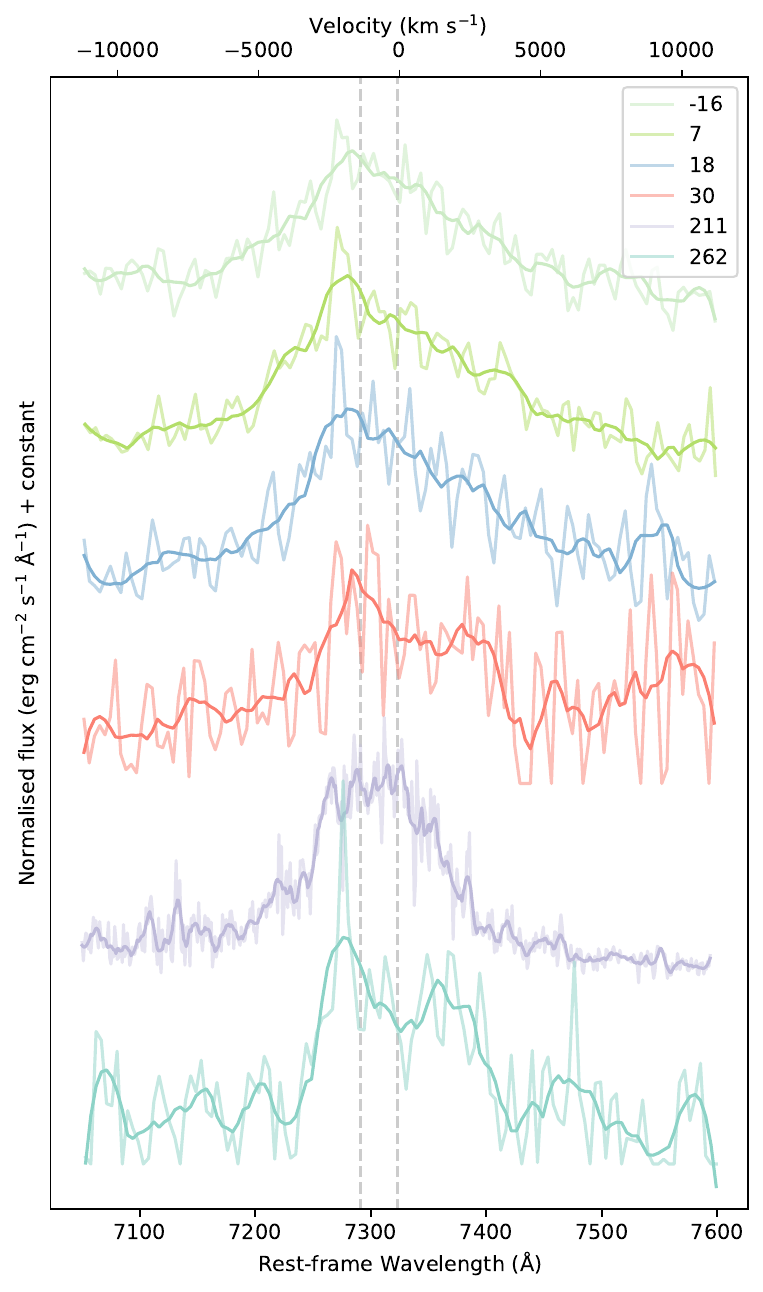}
    \caption{Evolution of the 7300\,$\Angstrom$ line centered on [O II] $\lambda\lambda$7320,7330 with darker lines indicating smoothed spectra and lighter indicating unsmoothed spectra. Dashed vertical lines represent [Ca II] $\lambda\lambda$7291,7323.}
    \label{fig:7300 line}
\end{figure}

\begin{figure*}
	\includegraphics[width=2\columnwidth]{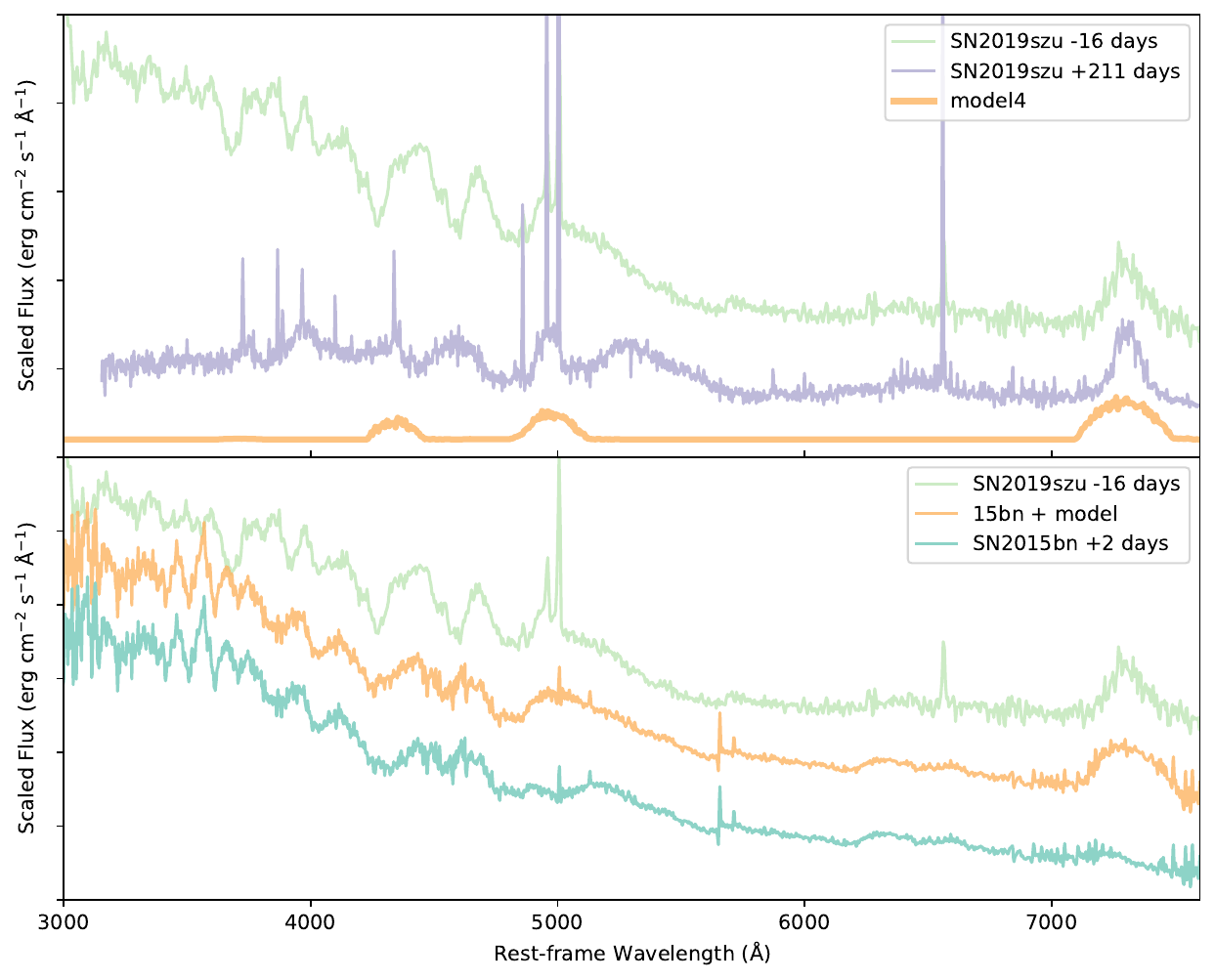}
    \caption{Comparisons of the SN2019szu with model spectra. \textbf{Top:} Photospeheric and nebular spectra of SN2019szu compared with a scaled version of the 104\_400 Cburn model from \citet{Jerkstrand2017b}. \textbf{Bottom:} The photospheric spectrum of SN2015bn near maximum light, and a composite model where the 104\_400 Cburn model has been added to this spectrum.  The SN2015bn spectrum is used to represent a characteristic SLSN spectrum and the model spectrum is used to represent a separate emitting region producing the forbidden emission lines.} This is compared with a photospheric spectrum of SN2019szu.
    \label{fig:composite}
\end{figure*}

\begin{figure*}
	\includegraphics[width=2\columnwidth]{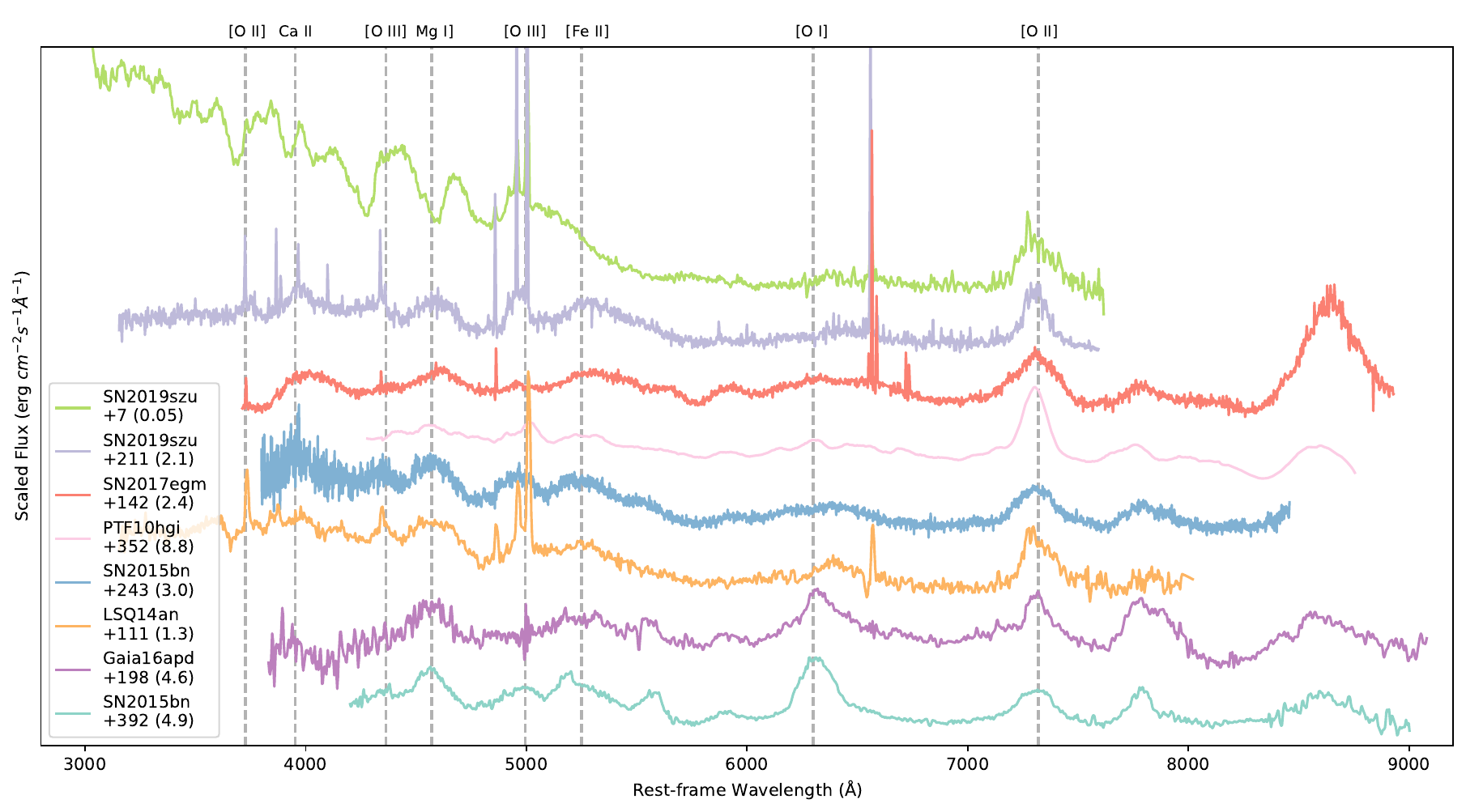}
    \caption{Comparison spectra of SN2019szu compared to a selection of other late time, nebular spectra of SLSNe showing the 7300\,$\Angstrom$ feature. Vertical dashed lines indicate positions of [O III] $\lambda$4363, [O III] $\lambda\lambda$4959,5007, [O I] $\lambda\lambda$6300,6364, and [O II] $\lambda\lambda$7320,7330. Phases given with respect to time of maximum light. Numbers in brackets represent the phase divided by exponential decline timescale for a more direct comparison of phase. Spectra are plotted by relative strength of [O I] and the emission line ~7300\,$\Angstrom$.}
    \label{fig:late time comp}
\end{figure*}

\subsection{Spectra}
\label{sec:spectra}
Figure \ref{fig:spec ev} shows the evolution of the spectra of SN2019szu, starting at $-$30 days pre-peak, up to 262 days post peak. The first spectrum shows a featureless blue continuum and although quite noisy, the narrow [O III] doublet at $\lambda$4959, and $\lambda$5007 from the host galaxy is visible. Later spectra were obtained using NTT and MMT, a full breakdown of this is given in Table \ref{tab:spectra}. The transition between a photospheric spectrum and a nebular spectrum occurs in the gap between the spectra at +31 and +211 days. Unfortunately this transition was not observed due to constraints from the Sun for ground-based telescopes.

\subsubsection{Photospheric Spectra}
\label{sec:early spec}
The `early' spectra here are defined as those taken before the break and showing photospheric absorption lines. In this case they run from $-$30 days to 30 days post peak in the rest frame. Narrow host-galaxy lines are seen with the [O III] emission lines at $\lambda\lambda$4959, 5007, and $\rm{H\alpha}$ at $\lambda$6563. The characteristic steep blue continuum and O II absorption lines associated with SLSNe are apparent in the 3000-5000\,$\Angstrom$ range, as well as typical Fe II and Fe III lines around 5000\,$\Angstrom$ blended with the O II absorption lines. Figure \ref{fig:labelled} shows the most prominent lines that can be seen in the spectra. An approximate SN velocity can be determined by measuring the blueshift of the O II absorption lines \citep{Gal-Yam2019a} yielding an ejecta velocity of $v_{\rm{ej}} \sim 4500 \rm{\,km\,s}^{-1}$. \citet{Chen2023b} found in their sample of events that the median O II derived velocity was 9700\,\kms around time of maximum light. SN2019szu has a considerably slower velocty but still within the range measured by their sample.

These spectra also include a broad emission line at 7300\,$\Angstrom$, an unusual feature at such an early phase for any SN, though a similar feature has been seen as early as $\sim$50 days post-peak in a handful of slowly evolving SLSNe \citep{Gal-Yam2009, Nicholl2016b, Inserra2017}. There is only one other event SN2018ibb, in which this feature has been observed around peak \citep{Schulze2023}. We will discuss this in detail in Section \ref{sec:7300 line}.

In Figure \ref{fig:spec ev} one can clearly see a lack of evolution in SN2019szu between $-$16 and +30 days. Other SLSNe have shown similar periods with minimal spectral evolution around maximum light; for example, SN2015bn maintained a constant spectrum dominated by O II and Fe III between at least $-$27 and $+$7 days \citep{Nicholl2016b}, but its spectrum then cooled and evolved quickly between 7 and 30 days. SN2019szu shows little evolution until at least 30 days, and if it did undergo any period of rapid cooling this was unfortunately unobserved while the object was in solar conjunction. This could be due to all of these spectra being obtained during the plateau at peak. The lack of line velocity evolution over this time also may favours the magnetar central engine for this event \citep{Mazzali2016}. The only feature that appears to change in this time period is the small bump that increases around 4340\,$\Angstrom$. This wavelength is consistent with [O III] $\lambda$4363, indicated by the first dashed grey line on Figure \ref{fig:spec ev}. This line has not been identified previously in such an early spectrum of a SLSN, and we explore this possible identification in Section \ref{sec:7300 line}. Any variation in [O III] $\lambda\lambda$4959,5007 over this time frame is hard to determine due to its blend with Fe II lines. There is also an unidentified emission line that varies in flux around 3850\,$\Angstrom$. If this was caused by Ca II H $\And$ K, it would require a blueshift of 7500\,\kms, and if it was caused by [O II] $\lambda$3727 it would require a redshift of 10500\,\kms, both of which seem unlikely due to the lack of other features with similar blue/redshifts.

\subsubsection{Continuum Shape}
\label{sec:continuum}

The shape of the continuum in the early time spectra shows a relatively flat region between 5500-7000\,$\Angstrom$ combined with a steep blue shape below $\sim$5500\,$\Angstrom$. We attempted to fit the continuum with a single blackbody (Figure \ref{fig:best bb fit}) and with the sum of two blackbodies of independent temperature. Our best fits indicate a singular source at $\sim$13000\,K which captures most of the shape of the flat region. Adding a second component for a double blackbody did not improve this fit and in fact preferred assigning both components the same temperature. Moreover, the continuum shape between 5500-7000\,$\Angstrom$ does not resemble other SLSNe at a similar epoch, most of which can be well approximated by a simple blackbody. 

A possible explanation for this could be an additional pseudo-continuum. In this case interaction with CSM produces a forest of narrow Fe II lines blended together, bluewards of $\sim$5500 \citep{Inserra2016b}. This has been observed in various types of interacting supernovae including Type Ia-CSM and Type Ibn such as SN2014av and SN2006jc \citep{Pastorello2016}. Figure \ref{fig:best bb fit} shows a spectrum for SN2021csp, a SN Icn that has been argued to have exploded within H and He-poor CSM. This event shows a strong pseudo-continuum bluewards of $\sim5000\,\Angstrom$ attributed to a forest of Fe lines produced due to interaction with the surrounding medium \citep{Perley2022, Fraser2021}. Figure \ref{fig:best bb fit} shows a composite spectrum created by summing a blackbody at 20000\,K to an arbitrarily scaled spectrum of SN2021csp. A hotter blackbody was needed as the best fit at 13000\,K under predicted in the bluer wavelengths. This approximately recreates the unique continuum shape of SN2019szu, with a flat red region combined with the steep continuum in the blue. In reality, the component originating from the SN ejecta is more complicated than the simple blackbody used here (e.g. the SN spectrum contains O II absorption lines). However we use this composite model purely to show we can achieve a similar continuum shape to SN2019szu with an additional interaction component. 

The pseudo-continuum in SN2019szu is apparent in the spectrum obtained at -16 days relative to peak. This suggests the SN was already interacting with CSM by this phase.

\subsubsection{The 7300\,$\Angstrom$ Line}
\label{sec:7300 line}

SN2019szu shows a prominent broad emission line at $\sim$7300\,$\Angstrom$ throughout its photospheric phase. While this line has been observed in some SLSNe in the late photospheric phase, it is already apparent in the earliest spectrum of SN2019szu that covers this wavelength range, meaning it is present at least 16 days before maximum light (Figure \ref{fig:7300 line}). The line itself does not evolve much over the course of our observations, with a similar full width at half maximum (FWHM) of $\sim7000$\,\kms\ during the photospheric phase. This velocity drops to $\sim5000$\,\kms\ FWHM in the late-time spectra which could be caused by a velocity gradient or change in optical depth. The total line flux stays relatively consistent during the early spectra at around $2\times10^{-14}$ erg cm$^{-2}$ s$^{-1}$ and decreases to $3\times10^{-15}$ erg cm$^{-2}$ s$^{-1}$ in the late-time spectra. 

Other slow-evolving events such as SN2007bi, PTF12dam, SN2015bn, and LSQ14an have shown early emission of forbidden and semi forbidden lines ranging from 50-70 days post peak \citep{Gal-Yam2009, Young2010, Nicholl2013, Nicholl2016b, Inserra2017}. One of the slowest events SN2018ibb displayed forbidden emission lines even earlier, apparent in its earliest spectrum at $-$1.4 days relative to peak \citep{Schulze2023}. Forbidden lines are formed when the radiative de-excitation dominates rather than collisional de-excitation. The conditions needed to form these forbidden emission lines are generally not seen until the SN reaches its nebular phase, when material is much more diffuse, temperatures are lower, and the energy deposition from the power source is lower \citep{Jerkstrand2017a}.

In previous SLSNe, the line at 7300\,$\Angstrom$ was usually identified as [Ca II] $\lambda\lambda$7291,7323. SN2018ibb is an interesting case as it is one the earliest examples of a SLSN showing nebular emission lines at 1.4 days before peak \citep{Schulze2023}. This was the earliest spectrum obtained of this object but still showed evidence for an emission line at 7300\,$\Angstrom$ that strengthened over time. The line profile shifted from a top-hat shape to bell shaped and also shifted by a few $\Angstrom$ redwards over the first 100 days. \citet{Schulze2023} attribute this to the profile originally displaying [Ca II] and slowly becoming dominated by [O II] $\lambda\lambda$7320,7330. In the case of LSQ14an, the line was strong in the earliest spectrum obtained 55 days after peak. \citet{Jerkstrand2017b} attribute this line to emission from [O II] $\lambda\lambda$7320,7330 as opposed to [Ca II] as identified in other SLSNe spectra, due to the strength of this line and the lack of [O I] $\lambda$6300 emission throughout the nebular phase, suggesting oxygen was primarily in higher ionisation states throughout the ejecta. \citet{Inserra2017} suggest this line is a combination of both [O II] and [Ca II] in order for the line to match the widths of other [O III] lines present. The line also appears to be slightly asymmetric with a small narrow peak on the blue side of the profile which is similar to the shape of the line in SN2019szu (Figure \ref{fig:7300 line}). This similar asymmetry could indicate the presence of [Ca II] in the SN2019szu spectra as the blue peak is close to the rest-frame wavelength of [Ca II] $\lambda$7291. However the low spectral resolution and signal-to-noise ratio makes it difficult using the line profile alone to determine to what extent [Ca II] or [O II] may be contributing. SN2007bi also showed a strong emission line at 7300\,$\Angstrom$ in its earliest spectrum at 50 days post-peak \citep{Gal-Yam2009}, though in this case it exhibited a more typical nebular phase dominated by [O I] $\lambda$6300.

Other effects could also play a role in the asymmetry seen for the 7300\,$\Angstrom$ line in SN2019szu. Continuous scattering can be caused by free electrons or dust resulting in a profile with a blueshifted peak with a longer red tail \citep{Jerkstrand2017a}. Considering the first epoch in Figure \ref{fig:7300 line}, the base velocity of the line is $\sim6000$\,\kms with a peak at $\sim-1500$\,\kms. This would put us in the regime where the electron scattering optical depth $\tau_{e}=2-3$ \citep{Jerkstrand2017a}. The evolution of the line then would be consistent with a declining $\tau_{e}$. In the nebular phase $\tau_{e} \lesssim 1$, leading to a diminished impact on the perceived blueshift of the line \citep{Jerkstrand2017a}. But semi-nebular lines can still emerge even with electron scattering depths of a few, and so the true velocity of the line may be slower. In some Type Ia SNe a similar effect of a blueshifted profile is thought to be caused by an aspherical explosion. \citet{Dong2018} show that in SN1991bg-like events, the central peak can be shifted off-center by $\sim$1000\,\kms. Alternatively the profile could be explained by the SN blocking the view of the fastest receeding material, thus resulting in a slight blueshift.

Although the 7300\,$\Angstrom$ line appears to be double peaked in the +30 day and +262 day spectra, we believe this is an artefact from the smoothing. This is apparent by looking at the line in the +211 day spectrum obtained using Binospec which has a better wavelength resolution and does not indicate a double peaked nature. However, double peaked features have been observed in SLSNe such as SN2018bsz which displayed a H$\alpha$ profile with a strong redshifted peak, and a weaker blueshifted peak. An asymmetric, disk-like CSM structure was used to explain how this profile could form \citep{Pursiainen2022}.

SN2019szu is the first SLSN that shows the 7300\,$\Angstrom$ line in a significantly pre-maximum spectrum, present alongside the characteristic O II absorption lines. At the early times seen in this event, the expected high radiation temperatures would imply that Ca II would be ionised and so lines from this species would not be observed. This is supported by the singly ionised oxygen lines evident in Figure \ref{fig:labelled}; O II has a higher ionisation potential than Ca II. This suggests that [O II] may be a more likely explanation for an emission line at 7300\,$\Angstrom$ appearing so early in the SN evolution.

We can test whether [O II] is a valid identification for this line by comparing to spectral models to predict other lines we expect to appear in these spectra. Nebular models from \citet{Jerkstrand2017b} predict emission from [O III] $\lambda\lambda$4959, 5007, and [O III] $\lambda$4363 to also appear if strong [O II] is present. Looking at Figure \ref{fig:labelled}, one can see evidence for these lines with the same blueshift as the [O II] emission line. This is even more apparent in Figure \ref{fig:composite}: adding the nebular model from \citet{Jerkstrand2017b} showing the [O II] and [O III] lines to a photospheric spectrum of SN2015bn greatly improves the match with SN2019szu in the areas with the [O II] and [O III] lines. It also recreates the profile seen around 5000\,\AA\ due to the blend of [O III] with Fe II/III. This method of combining photospheric and nebular spectra assumes the nebular component is transparent enough to not impede our view of the SN, and has been used before in the literature. In particular \citet{Ben-Ami2014} used this method for SN2010mb, a SN with evidence for interaction with large amounts of hydrogen-free CSM. With this in mind we identify the emission line at 7300\,$\Angstrom$ as [O II] $\lambda\lambda$7320,7330 with a net blueshift indicating a velocity $\sim 1500$\,\kms. We will explore the physical implications of this line in Section \ref{sec:CSM params}.

\subsubsection{Late Spectra}
\label{sec:late spec}

At late times the spectrum has transitioned into its nebular phase, defined by the broad emission lines and lack of thermal continuum. The spectra of SN2019szu at these phases are much more consistent with other SLSNe, as can be seen in Figure \ref{fig:late time comp}. Prominent lines (labelled in Figure \ref{fig:labelled}) include [O II] $\lambda$3727, [O II] $\lambda\lambda$7320,7330, Ca II $\lambda\lambda$3934,3969, [O III] $\lambda$4363, [O III] $\lambda\lambda$4959,5007, Mg I] $\lambda$4571, and a blend of iron lines around 5200\,$\Angstrom$ including  [Fe II] $\lambda$5250.

One notable difference is the lack of [O I] $\lambda$6300 emission in SN2019szu. This line is visible in most of the spectra of the other SLSNe at a similar phase and so reflects a high and persistent degree of oxygen ionisation in SN2019szu. In the sample of SLSN nebular spectra analysed by \citet{Nicholl2019c}, three out of 12 events showed evidence for weak [O I] $\lambda$6300 and with [O II] dominating the 7300\,$\Angstrom$ region. This could be due to runaway ionisation which is believed to occur in the magnetar central engine scenario for low ejecta masses and a high power pulsar wind nebula. The result of this is a sharp switch of the spectra from O I dominated to O II/O III dominated. In the O II/O III dominated space this leads to a suppression of [O I] $\lambda$6300 emission as seen in SN2019szu \citep{Jerkstrand2017b, Omand2023}. Alternatively, this line could emerge at later phases even for events with massive ejecta. SN2015bn and LSQ14an are modelled to have $\gtrsim$10\,M$_{\odot}$ of ejecta and show little [O I] emission at early on in their nebular phase (+243 and +111 days respectively). But both events show much stronger emission after 300-400 days \citep{Jerkstrand2017b, Nicholl2016c}. We do not have spectra for SN2019szu at such a late phase.

The [O II] and [O III] emission lines, previously blended with broad absorption lines of O II and Fe III, are now much more easily isolated, confirming our earlier identification of [O II] $\lambda\lambda$7320,7330.

\begin{figure}
	\includegraphics[width=\columnwidth]{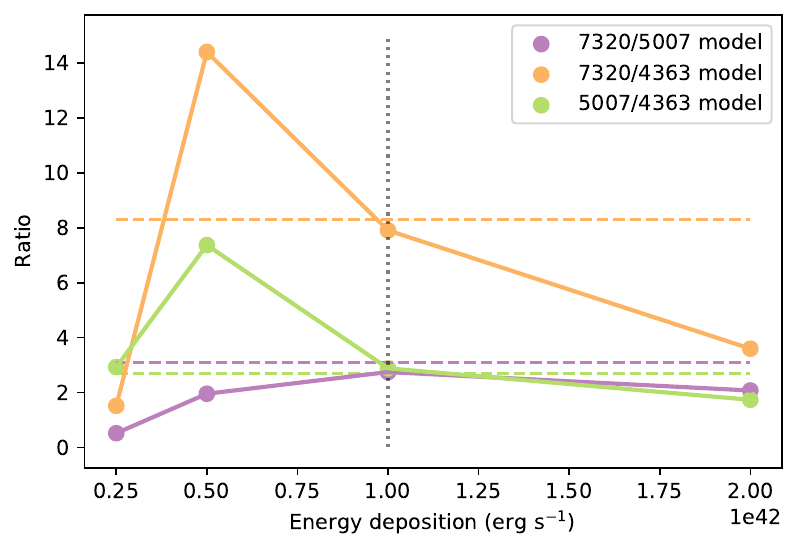}
    \caption{Line ratios of [O II] and [O III] lines in model Cburn spectrum with M=3\,M$_{\odot}$ and f=0.1 \citep{Jerkstrand2017b}. Dashed horizontal lines indicate line ratios found in the +262 day spectrum for an unknown energy deposition. Our data matches well with all the line ratios at $E_{\rm{dep}}=1\times10^{42}$\,erg\,s$^{-1}$.}
    \label{fig:line ratios}
\end{figure}

\begin{table}
	\centering
	\caption{SUMO models used. All models assume ejecta 400 days post explosion with a velocity of 8000\,\kms, and 100 randomly distributed spherical clumps. M is the mass ejected in each model, f is the filling factor, and E$_{\text{dep}}$ is the energy injected into the system.}
	\label{tab:model spectra}
	\begin{tabular}{lccr} 
		\hline \hline
		Model & M (M$_{\odot}$) & f & E$_{\text{dep}}$ (erg/s)\\
		\hline
            101\_400 & 3 & 0.1 & 2.5 $\times$ 10$^{41}$ \\
            102\_400 & 3 & 0.1 & 5 $\times$ 10$^{41}$ \\
            103\_400 & 3 & 0.1 & 1 $\times$ 10$^{42}$ \\
            104\_400 & 3 & 0.1 & 2 $\times$ 10$^{42}$ \\
            108\_400 & 3 & 0.01 & 2 $\times$ 10$^{42}$ \\
		\hline \hline
	\end{tabular}
\end{table}

\section{Modelling and Interpretation}
\label{sec:Models}

We have demonstrated that SN2019szu is a bright and slowly evolving SLSN, which displays surprising and persistent [O II] and [O III] emission lines even in early spectra obtained 16 days before maximum light. Our goal for this section is to determine the physical parameters of the oxygen emitting region and the ejecta overall, in order to understand how forbidden emission can arise at a time when the density in the ejecta is typically expected to be too high.

The typical electron density of expanding ejecta can be written as:

\begin{equation}
    \begin{split}
        n_e =& \enspace 2\times10^{9} \mu^{-1} 
        \left(\frac{M_{\rm ej}}{\rm{M}_\odot}\right)\, 
        \left(\frac{v_{\rm ej}}{3000\,\rm{km\,s}^{-1}}\right)^{-3}\, \\
        & \times \left(\frac{x_{e}}{0.1}\right)\,
        \left(\frac{t}{200\,{\rm days}}\right)^{-3}\, 
        \left(\frac{f}{0.1}\right)^{-1}\,
        \rm{cm}^{-3}
    \end{split}
\end{equation}

where $\mu$ is the mean atomic weight, $M_{\rm{ej}}$ and $v_{\rm{ej}}$ are the ejecta mass and velocity respectively, $x_e$ is the electron fraction, $t$ is the time from explosion, and f is the filling factor \citep{Jerkstrand2017a}. The critical density for the [O II] line is $n_{\rm crit}\approx10^7$\,cm$^{-3}$ \citep{Appenzeller1988}. Above this density, collisional de-excitation dominates rather than radiative de-excitation, therefore suppressing the formation of forbidden lines. Assuming the ejecta is mostly oxygen, with a velocity of $\sim10000$\,\kms, at 80 days after explosion this results in $n_{e} \sim 10^{10}\,(M_{\rm{ej}}/$M$_\odot$)\,cm$^{-3}$. For typical SLSN ejecta masses of around 10\,M$_\odot$ \citep{Blanchard2020}, this is several orders of magnitude greater than the [O II] critical density. This suggests that [O II] emission is more likely to arise from lower density material, presumably expelled before the first detection of SN2019szu. The early appearance of these lines requires material close to the SN site, therefore excluding material other than that originating from the SN or its progenitor. This argument still holds even if the 7300\,$\Angstrom$ line has been misidentified, since the critical density for [Ca II] is similar to that for [O II]. As discussed in Section \ref{sec:continuum}, the unusual shape in the spectral continuum may be attributed to interaction, providing further evidence for a (presumably H-poor) CSM close to the explosion site.

\subsection{Models for the Oxygen Emission Lines}
\label{sec:CSM params}

\citet{Jerkstrand2017b} investigated the emission from long-duration SLSNe during their nebular phase using models from their SUMO code. These models explore oxygen-rich compositions and single zones, as this allowed exploration of parameters agnostic to the powering mechanism \citep{Jerkstrand2017b}. Different compositions included a pure oxygen zone (pureO), and carbon burning ashes (Cburn). Here we choose the latter series to explore fully, as it is more physically motivated. Element abundances for this model were taken from the ONeMg zone in the \citet{Woosley2007} models assuming a star $M_{\rm ZAMS} = 25$\,M$_{\odot}$ collapsing into a supernova. The model assumes 100 randomly distributed spherical clumps with vacuum in between, at 400 days post explosion. The ejecta is also assumed to have been travelling at a constant velocity of 8000\,\kms. The energy deposited is assumed to be from high energy sources such as gamma rays. Each model had varying ejecta mass ($M_{\text{ej}}$), filling factor ($f$), and energy deposition ($E_{\text{dep}}$).

A small subset of models were chosen from the grid of parameters to explore based on the line strengths of [O I] $\lambda\lambda$6300, 6364 compared to [O II] $\lambda\lambda$7320, 7330, where any models with visible [O I] emission compared to [O II] were rejected. This is motivated by Figure \ref{fig:spec ev}, where these particular lines are isolated from other lines in the late-time spectra of SN2019szu, and the lack of [O I] emission is clear. In the models, a lack of [O I] can occur due to near-complete oxygen ionisation. A full breakdown of models passing our selection is given in Table \ref{tab:model spectra}. The 104\_400 model also creates [O III] lines at 4363\,$\Angstrom$, 4959\,$\Angstrom$, and 5007\,$\Angstrom$ which when scaled and added to an early spectrum of SN2015bn in Figure \ref{fig:composite} recreates the line profile seen in SN2019szu around 5000\,$\Angstrom$. The 4363\,$\Angstrom$ line also lines up with the emission feature seen in Figure \ref{fig:spec ev} that appears to increase in strength over time. This provides solid evidence that these additional features can be explained by oxygen rich material. All of these oxygen lines also have a similar blueshift derived from the peak of the [O II] $\lambda\lambda$7320,7330 and [O III] $\lambda$4363 lines, which indicates a velocity of $v=1500$\,\kms.

We now seek to determine the heating rate and the density of oxygen needed to explain the [O II] line luminosity in SN2019szu. In the sample of four models, the luminosity of the 7300\,$\Angstrom$ line is roughly proportional to the energy deposited. From this we can estimate the energy deposition that gives the correct line luminosities for the SN2019szu spectra. The luminosity of the [O II] line in SN2019szu is $\sim 1 \times 10^{42}$\,erg\,s$^{-1}$ at early times and $\sim 4 \times 10^{41}$\,erg\,s$^{-1}$ in our nebular spectra, which would require $E_{\text{dep, early}} \sim 4 \times 10^{42}$\,erg\,s$^{-1}$ and $E_{\text{dep, late}} \sim 1 \times 10^{42}$\,erg\,s$^{-1}$ assuming the [O II] luminosity to be directly proportional to deposition. The required deposition at early times is also outside the range explored by \citet{Jerkstrand2017b} but only by a factor of 2. Extrapolating to higher energies is not trivial, as changing the deposition does not only affect the line luminosities; it will also influence the temperature and ionisation of the ejecta, and hence the line ratios. In order to keep the ionisation state constant for different energies and masses, the following relation must hold:

\begin{equation}
    \frac{\gamma}{\alpha n_{e}} = \text{constant},
    \label{eq:gammaalphanO}
\end{equation}

where $\alpha$ is the recombination rate, and $n_{e}$ is the electron number density. $\gamma$ is the ionisation rate per particle, defined as:

\begin{equation}
    \gamma = \frac{e_{\text{dep}} \: x_{\rm{ion,OII}}}{ \chi \: n_{\rm{OII}}}
    \label{eq:gamma}
\end{equation}
 
Here $x_{\rm{ion,OII}}$ is the fraction of deposited energy used in ionising O II to O III, $e_{\text{dep}}$ is the energy deposition per unit volume, and $\chi$ is the ionisation potential.  We can approximate this as $x_{\rm{ion,OII}} \sim x_{\rm{ion}}x_{\rm{OII}}$, where $x_{\rm{OII}}$ is the fraction of O II in the gas. We assume $n_{e}$ to be proportional to the oxygen number density $n_{O}$, which is then proportional to the density of the material $\rho$ for our sample of models. This is motivated by the grid of models which show a ratio of $n_{e}/n_{O}\sim1-1.3$ for the subset used \citep{Jerkstrand2017b}. If $\alpha$ is not too strong a function of temperature, we can assume it will be constant for all of these models with the same composition. This leads to the relation 

\begin{equation}
    \frac{E_{\rm{dep}} x_{\rm{ion}}}{\rho^{2}} = \text{constant}
    \label{eq:edep}
\end{equation}

which can then be used to calculate the density of the oxygen-emitting material needed in order to maintain a constant ionisation fraction for a given energy deposition. The fraction of energy going into ionisations, $x_{\rm{ion}}$, decreases with higher energy depositions, but the SUMO models show that this function varies very slowly and we may assume it to be constant in the regime we are working in.

To use the relation in Equation \ref{eq:edep}, we first identified models that produced lines consistent with SN2019szu. This was done by comparing the line ratios of [O II] $\lambda\lambda$7320, 7330, [O III] $\lambda$4363, and [O III] $\lambda\lambda$4959, 5007 in the model series to our observed spectra. This is shown in Figure \ref{fig:line ratios} for the spectrum at +262 days. This method was only applicable to the final two nebular spectra in Table \ref{tab:spectra}, as the lines were not isolated enough in earlier spectra. Both spectra indicated $E_{\text{dep}} \simeq 1\times 10^{42}$ erg s$^{-1}$ provided roughly the correct ionisation balance for a 3\,M$_{\odot}$ model with a filling factor $f=0.1$. This model has $\rho_{\rm{model}} = 6.7 \times 10^{-16}$\,g\,cm$^{-3}$. We verified that applying a significant host galaxy extinction (Section \ref{sec:host}) did not change the line ratios significantly and so we disregarded this effect.

We can thus scale the deposition to match the observed spectra, and determine the corresponding density to match the inferred ionisation state. Using $E_{\text{dep, early}}$ and $E_{\text{dep, late}}$, this resulted in $\rho_{\rm{early}} \sim 1\times 10^{-15} \rm{\,g\,cm}^{-3}$ and $\rho_{\rm{late}} \sim 7\times 10^{-16} \rm{\,g\,cm}^{-3}$. However, as the ionisation state cannot be constrained directly in the early spectra, the early measurements assume a constant ionisation state over the course of the SN lifetime, and so $\rho_{\rm{early}}$ should be taken as a rough guide only. This density will be used further in Section \ref{sec:timescales} to constrain the CSM parameters. We also note that our measurement of the late time [O II] $\lambda\lambda$7320, 7330 line may be contaminated by nebular emission from the SN ejecta. Based on Equation \ref{eq:edep}, a small change in $E_{\rm{dep}}$ would result in an even smaller change in the density as it scales with $\sqrt{E_{\rm{dep}}}$, and so this calculation is not very sensitive to contamination from the ejecta. We can use the early time energy estimate as a check as we would not expect to see contamination during the photospheric stage. We can see that the density estimates at early and late times are within a factor of a few from each other.  Another check is the line flux remaining relatively constant over the course of its evolution and so we believe the line flux at times is dominated by CSM interaction.

\begin{figure*}
	\includegraphics[width=2\columnwidth]{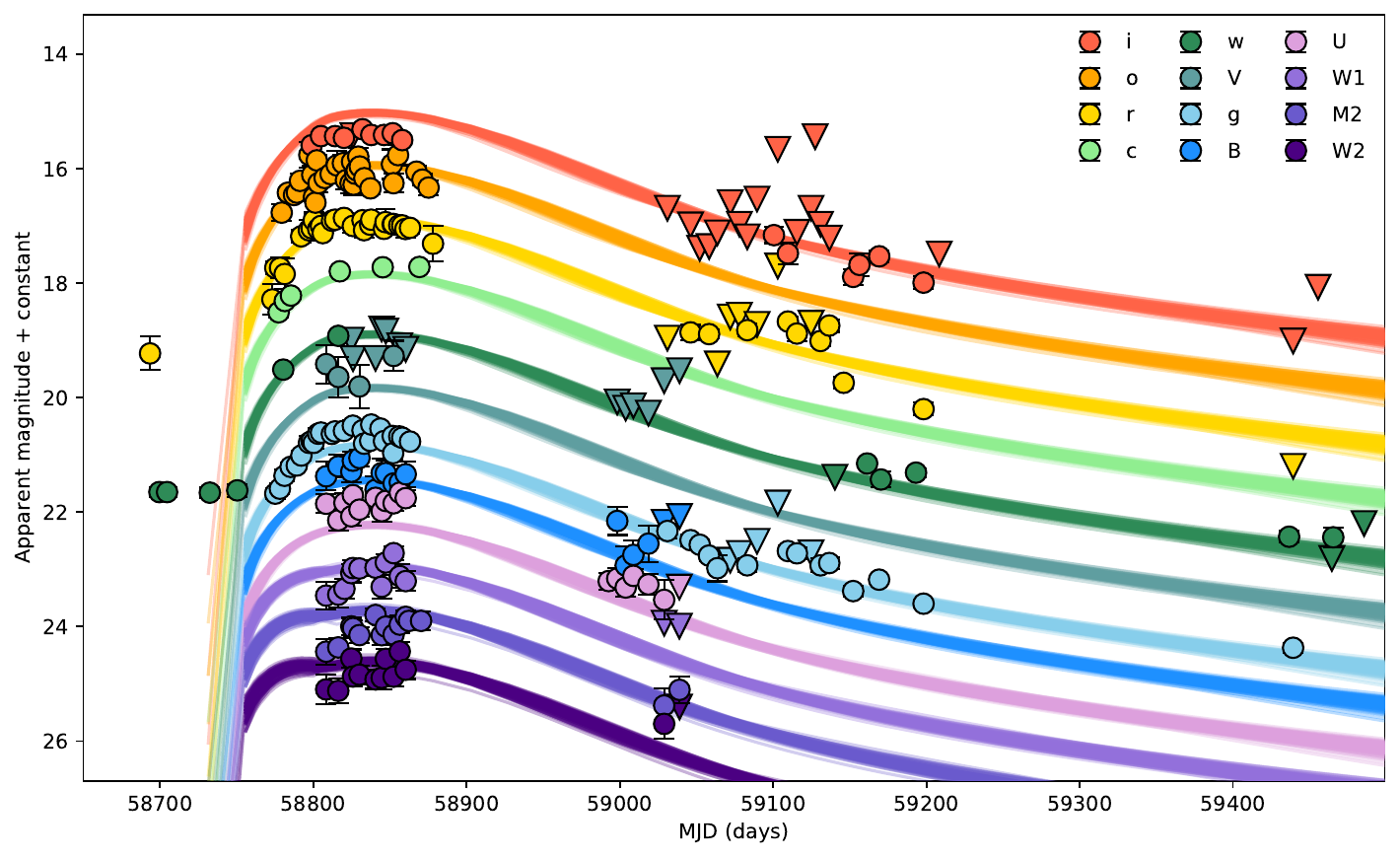}
    \caption{Magnetar model fits to SN2019szu using \textsc{mosfit}. Pre-explosion data is plotted but was excluded from the fits. Upper limits are indicated via inverted triangles and band offsets for display are: uvw2+5.5; uvm2+5; uvw1+4.5; U+4; B+2.5; g+2; V+1; c+0; w-1; r-2; o-3; i-4.}
    \label{fig:mosfit models}
\end{figure*}

\begin{figure*}
	\includegraphics[width=2\columnwidth]{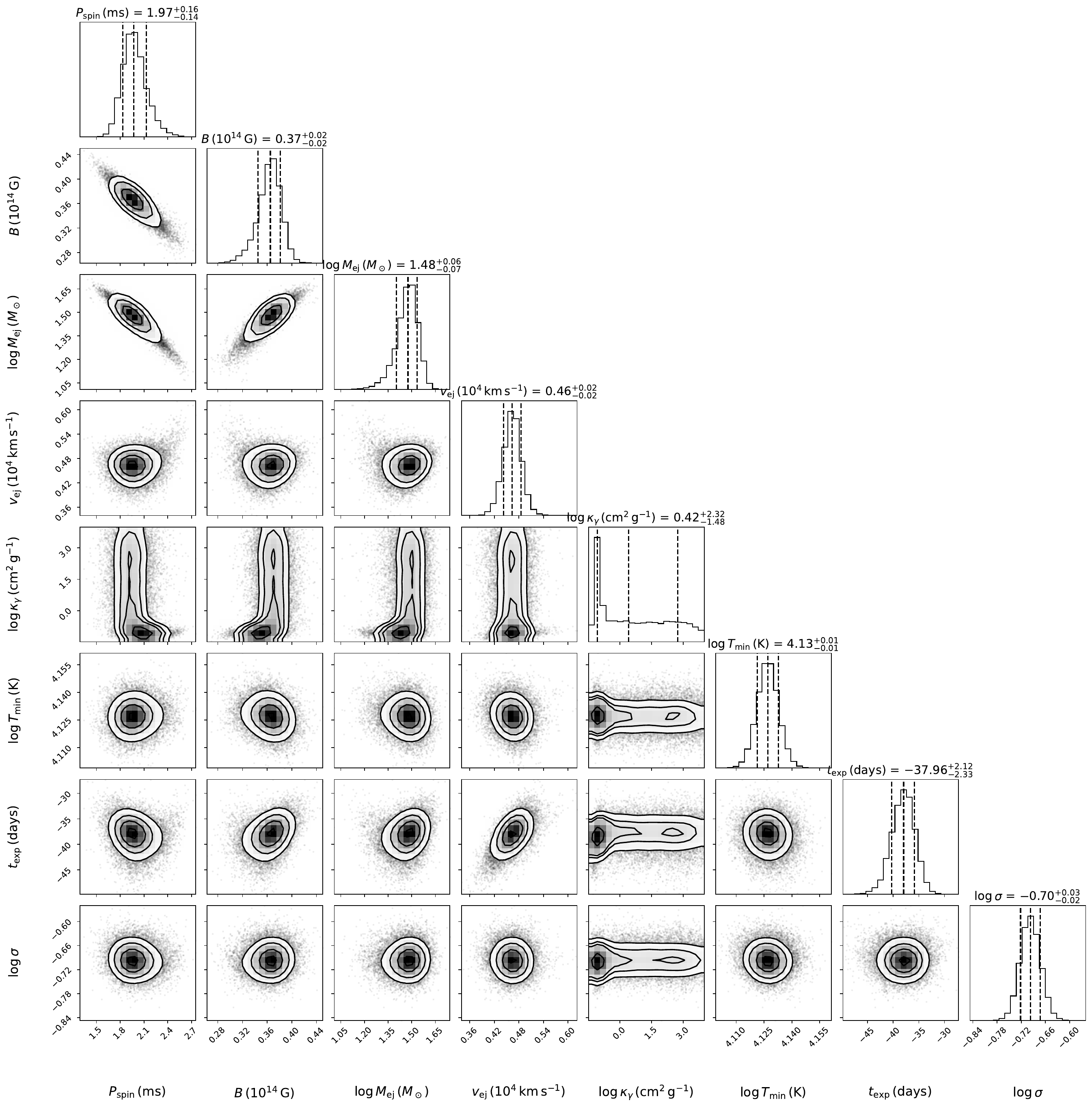}
    \caption{Posteriors for magnetar model fit to SN2019szu excluding pre-explosion data. Medians and 1$\sigma$ ranges are labeled.}
    \label{fig:mosfit corner}
\end{figure*}

\subsection{Magnetar Model and Ejecta Mass Estimate}
\label{sec:magentar models}

Modelling the light curve allows us to estimate the mass and velocity of the SN ejecta. Semi-analytic magnetar-powered models have been extensively tested in the literature and shown to able to reproduce the light curves of most observed SLSNe \citep[e.g.][]{Inserra2013b, Nicholl2017a, Chen2023a}. Although we have found evidence for CSM interaction in this event through the material needed to produce the 7300\,$\Angstrom$ line, the low density of this material suggests that it is likely not sufficient to power the full luminosity at the bright peak of the light curve, so multiple power sources may be at play. \citet{Nicholl2015a} modelled 24 SLSNe light curves using formulae detailed by \citet{Chatzopoulos2012}, and showed that densities of $10^{-12}$\,g\,cm$^{-3}$ were needed to match the rise-decline relation seen in SLSN light curves. Although these analytic CSM models are widely used \citep{Chatzopoulos2012}, the complicated geometry of the interaction and additional flexibility from separate ejecta and CSM density profiles can make the results difficult to interpret, and lead to mass estimates that are quite discrepant with hydrodynamical models \citep[e.g.][]{Sorokina2016,Nicholl2020,Suzuki2021}. We therefore restrict our mass estimates to the magnetar model only, but note that the additional presence of CSM introduces an additional systematic uncertainty. {We quantify the extent of the CSM contribution in Section \ref{sec:discussion}.}

To constrain the magnetar parameters that would be needed to power SN2019szu, and to estimate the ejecta mass and velocity, we employed the Modular Open Source Fitter for Transients (\textsc{mosfit}) \citep{Guillochon2018,Nicholl2017a}. \textsc{mosfit} is a fully Bayesian code that fits physical models to the multi-band light curves of transients. The plateau was excluded from this fitting as it is believed to be a result of pre-explosion activity and the magnetar model would not be able to fit such a feature. For the magnetar model the assumption is the energy input at time $t$ is given by:

\begin{equation}
    F_{\text{mag}}(t) = \frac{E_{\text{mag}}}{t_{\text{mag}}} \frac{1}{(1+t/t_{\text{mag}})^{2}}
    \label{eq:Fmag}
\end{equation}

\begin{equation}
    E_{\text{mag}} \propto P^{-2}
    \label{eq:E approx}
\end{equation}

\begin{equation}
    t_{\text{mag}} \propto P^{2} B_{\perp}^{-2} \text{s}
    \label{eq:t approx}
\end{equation}

Where $E_{\text{mag}}$ is the rotational energy of the magnetar, $t_{\text{mag}}$ is the timescale it spins down on. $P$ is the spin period of the magnetar and $B_{\perp}$ is the perpendicular component of the magnetic field to the spin axis.

The default \textsc{mosfit} priors were edited to account for the specifics of this event. In initial fits the value of A$_{\rm{V,host}}$ railed against the upper end of the default prior, with A$_{\rm{V,host}}$ approaching 0.5 mags, inconsistent with a dwarf host galaxy. The parameter fit in the models for this is the hydrogen column density ($n_{\rm{H}}$), which is related to A$_{\rm{V,host}}$ by $A_{\rm{V}} = n_{\rm{H}} / 1.8\times 10^{21}$. We therefore fixed $n_{\rm{H}}$ to 10$^{16}\,\rm{cm}^{-2}$ to ensure that an unrealistic extinction did not bias our fits. 

The prior for scaling velocity ($v_{\text{ej}}$) was set to be a Gaussian distribution with mean = 5000\,\kms\ and standard deviation = 3000\,\kms, based on the blueshift of the O II lines in the photospheric spectra as described in Section \ref{sec:early spec}. This approximation can be made for SLSNe around peak as the O II lines form in a region close to the photosphere. The shallow nature of the photosphere means it can be described using a single velocity \citep{Gal-Yam2019a}, which can be used as a proxy for the ejecta velocity. The broad Gaussian also allows for a more typical SLSN velocity $\sim 10,000$\,\kms\ as the lines may not have formed in the same location as the photosphere, and hence could have differing velocities. The prior on the minimum temperature ($T_{\text{min}}$) at late times was adjusted to cover a broader range from 10$^{3}$ to 10$^{5}$\,K with a flat distribution in log space, the physical motivation being the unusual behaviour of the colour temperature of SN2019szu (Section \ref{sec:colour}). The prior for the magnetic field ($B$) was expanded to include lower values from 10$^{12}$\,G to 10$^{15}$\,G to allow for the slow evolution of SN2019szu. The optical opacity $\kappa$ was fixed to 0.15 g cm$^{-2}$, the median value found in the population study by \citet{Nicholl2017a}.

Model fits are shown in Figure \ref{fig:mosfit models}, and the derived posteriors are given in Figure \ref{fig:mosfit corner}. Overall the model fits cannot fully capture the the bumps and wiggles in the light curves which would require more nuanced models. It also under predicts the luminosity in some of the bluer bands such as $V$, $g$, and $U$, whilst over predicting the $i$ band luminosity at peak, and the $r$ band at late times. This perhaps is a reflection of the unusual shape of the SED for this event. The magnetic field $B=0.37^{+0.02}_{-0.02} \times 10^{14}$\,G is at the low end for the population studied by \citet{Nicholl2017a} ($B = 0.8^{+1.1}_{-0.6}\times10^{14}$\,G), which is perhaps unsurprising given the slow decline together with Equation \ref{eq:t approx}. A lower value of $B$ results in energy deposition over a longer time frame, leading to a longer lived SLSN. We can also see that the spin period found for SN2019szu ($P=1.97^{+0.16}_{-0.14}$\,ms) is within the 1$\sigma$ range of the median value found \citet{Nicholl2017a} ($P=2.4^{+1.6}_{-1.2}$\,ms). The spin period sets the luminosity scale, as $F_{\rm{mag}}(t) \propto E_{\rm{mag}}/t_{\rm{mag}} \sim \rm{P}^{-4}$. As SN2019szu is a comparable luminosity to other SLSNe this is unsurprising.  

Our estimated ejected mass ($M_{\text{ej}}=30.2^{+4.5}_{-4.5}$) is also on the upper end of the general population, with a median and 1$\sigma$ distribution of $M_{\text{ej, median}} = 4.8^{+8.1}_{-2.6}$ \citep{Nicholl2017a}.  A larger population study by \citet{Blanchard2020} found this distribution of masses spanned $3.6-40 \,\rm{M}_{\odot}$ for the general population and so SN2019szu also lies on the upper end of this. The mass estimate is derived from the light curve width and diffusion timescale, and is not as dependent on the type of model chosen provided the power source is internal to the ejecta (e.g. $^{56}$Ni decay or magnetar engine). An interaction model is more complex, as the CSM can provide additional diffusive mass. However, since we have calculated the density of the CSM to be low relative to typical models for SLSNe, we would expect that the diffusion time is still dominated by the ejecta. For this reason, we expect our estimate of the total mass should be the right order of magnitude, even if interaction with the CSM is also contributing to the light curve. In SN2019szu, the large mass creates a longer diffusion time to contribute to the slow evolution. \textsc{mosfit} also provides an estimated explosion time relative to the first data point of $t \sim -37.96^{+2.12}_{-2.33}$\,days. Combining this with our time of maximum light, we estimate a total rise time of $\approx$82 days.

\subsection{Constraints on Pre-explosion Mass Loss}
\label{sec:timescales}

From previous analysis in Section \ref{sec:Models}, we know that SN2019szu has a region of low density, oxygen rich material needed to produce the forbidden oxygen lines at early times. The density of this material was calculated to be $\rho \sim 10^{-15}\,\rm{g\,cm}^{-3}$ in Section \ref{sec:CSM params}. This density is too low to be from the SN itself and so this region must be comprised of pre-expelled circumstellar material.

Using the velocities of the CSM and ejecta, we can constrain when this material was ejected before explosion, such that the ejecta catches the CSM prior to the first NTT spectrum. The width of the [O II] $\lambda\lambda 7320,7330$ line did not decrease much over time, changing from 7000\,\kms to 5000\,\kms over the course of $\sim$300 days. Throughout this period, it maintains a constant blueshift of $\approx1500$\,\kms. We assume that this blueshift corresponds to the CSM velocity and explore the consequences of this, though we acknowledge that it is also possible to achieve a blueshifted profile with electron scattering \citep{Jerkstrand2017a}.

The \textsc{mosfit} model provides a measurement for the scaling velocity $v_{\text{ej}} = 4700$\,\kms, which is consistent with the SN velocity estimated from the O II absorption lines measured in Section \ref{sec:early spec}. Another velocity estimate comes from the increase in radius (Figure \ref{fig:bb fits}). Assuming a $R=vt$ relation gives $v_{\text{ej}} = 1700$\,\kms over the first 30 days which is similar to the velocity estimates from the [O II] emission lines. This could suggest that at early times there is a photosphere within the CSM, however it is important to remember the limitations of the SED fitting used to produce these radii.

The presence of the [O II] emission line at $\approx$66 days from the estimated explosion date indicates that the SN ejecta has already caught up with the CSM at this phase, giving an upper limit on the CSM radius of $2.6 \times 10^{15}$\,cm. This places an upper limit on the ejection time which must be more recent than 200 days before the peak, or $\sim$120 days before explosion, for a CSM velocity of $v_{\text{CSM}} \sim 1500$\,\kms\ and SN velocity $v_{\text{SN}} \sim 4500$\,\kms. Mass ejection on a timescale of only months before the explosion is also supported by the precursor plateau in the light curve.

Using the estimated velocity and expansion time to determine the CSM radius at different times, we can also estimate the mass of emitting CSM. Using the approximate radius $\approx 2.6 \times 10^{15}$\,cm at the light curve peak results in $M_{\rm{early}}\sim0.1$\,M$_{\odot}$, whereas in the late spectra at $\approx$250 days post peak we find $M_{\rm{late}}\sim0.25$\,M$_{\odot}$ based on the densities found in Section \ref{sec:CSM params} and assuming spherical, uniform CSM. However, contamination from nebular lines at late times could result in a slight increase in $M_{\rm{late}}$. Both masses being similar suggest that the the SN had interacted with the majority of this CSM by the time of our spectral observations. However if this slight increase in apparent mass is real, it could point to a thicker shell, meaning the SN is able to interact with more material as it expands.  It is also important to note that the ionisation state derived from the late-time spectra may not strictly hold at earlier times.

Using our CSM mass estimate of around $0.25$\,M$_{\odot}$ and the velocity of this material we can calculate the total kinetic energy (KE). We obtain a value of KE $\sim 10^{49}$\,erg adopting the velocity from the [O II] and [O III] emission lines. This is a small fraction of the total energy obtained by integrating the bolometric light curve of E$\sim 3\times 10^{51}$ erg and so only a small fraction of the total energy would be released via these line emissions.

\section{Discussion}
\label{sec:discussion}

\subsection{The progenitor of SN2019szu}

We have found that the early emission lines in SN2019szu require 0.25\,M$_{\odot}$ of CSM ejected less than 200 days before maximum light. Explaining the mass ejected from the star in such a short amount of time requires creative explanations, as stellar wind mass loss rates, even for Wolf Rayet stars, are too low to explain the entirety of this CSM \citep{Mauron2011, Sander2022}. Typical mass loss rates for these types of stars range between $(0.2-10)\times 10^{-5}\,\rm{M_{\odot}}\, \rm{yr}^{-1}$ \citep{Nugis1998, Crowther2007}, and so more explosive mechanisms are required to explain this amount of CSM. A consistent picture must also be able to reproduce the luminous precursor plateau in the light curve.

Eruptions from luminous blue variable (LBV) stars have been proposed as a way of generating CSM in the context of powering Type II SLSNe. Giant LBV eruptions like that of $\eta$ Carinae can have mass loss rates $\gtrsim$0.5\,M$_{\odot}$yr$^{-1}$ \citep{Smith2003}, which could be sufficient to produce the quantity of CSM surrounding SN2019szu. However, these events are too faint to explain the pre-explosion activity seen in SN2019szu, with events such as $\eta$ Carinae reaching a maximum bolometric absolute magnitude of $M_{\rm{bol}}\sim-14$\,mag \citep{Smith2011}. SN2009ip was an event that bridged the gap between LBVs and SNe. Initially recognised as an LBV after a series of outbursts, the event transitioned into a SN-II reaching a maximum absolute magnitude $M_{r}=-17.5$\,mag \citep{Mauerhan2013}. Its spectra displayed broad Balmer lines with P-Cygni profiles characteristic of SNe-II. As LBV outbursts tend to occur in luminous stars in the blue
supergiant phase rather than in stars which are stripped cores, 
this explanation makes it very difficult to explain the lack of hydrogen
 and helium emission from the oxygen emitting region of SN2019szu.
However, we note that at least one precursor outburst has been
seen in a Ibn \citep{Pastorello2007}. 

Another possible explanation for the origin of this CSM is pulsational pair instability (PPI) ejections. \citet{Woosley2017} explores the evolution of stars thought to undergo PPI and describes temperature and luminosity models for these events. Less massive cores lead to smaller PPI shell masses at times closer to core collapse. In the models with He core masses between 48-52\,M$_{\odot}$ (CO cores of $\gtrsim40$\,M$_{\odot}$), the shells of material ejected in each pulse collide with one another and produce sustained luminosities above $\sim10^{43}$\,erg\,s$^{-1}$ from 40 days before explosion. This could provide an explanation for the pre-explosion plateau observed in SN2019szu and would be the first direct observation of PPI ejections if true. This subset of models also describe typical timescales for the onset of the first PPI ejections before core collapse of $>3.6 \times10^{6}$\,s. This is also consistent with the timescales obtained from the 7300\,$\Angstrom$ line, which constrained the time of mass ejection to within the last 120 days before time of maximum light.

However, this subset of models produce ejected shell masses in the region of 6-8\,M$_{\odot}$. This is significantly more massive than the $\approx0.25$\,M$_{\odot}$ of CSM responsible for the oxygen emission lines in SN2019szu, which would instead be consistent with $\approx30$\,M$_\odot$ cores. However, at lower masses the PPI eruptions occur only in the final days before explosion, inconsistent with the light curve plateau. {Alternatively, the discrepancy in CSM mass could be explained by the loss of the He layers before the onset of PPI, resulting in less massive shells of CO material. In fact, this would likely favour the scenario where the outermost shells of the PPI ejections produce the forbidden oxygen emission lines.} We suggest that a stripped $\sim40$\,M$_{\odot}$ CO core is more consistent with the observations, and the low mass, low density material producing forbidden lines is the outermost material from the earliest pulse. This assumption is supported by \citet{Renzo2020}, who show that the density of the PPI ejecta decreases as you move radially outwards for a 50\,M$_\odot$ He core ($\sim 40$\,M$_\odot$ CO core). Their work also suggests that these pulses may have velocities of a few thousand \kms, in alignment with our measured velocity from the 7300\,$\Angstrom$ line. Later successive pulses interact with each other to produce the plateau \citep{Woosley2017}, and the ejecta interacting with these shells could be the origin of the pseudo-continuum. We also note that many factors could affect our calculated CSM mass, such as the clumpiness of the ejecta or how much of the material has been excited. We assumed spherical geometry and that the material is shock excited, but it could also be radiatively excited. Nonetheless, it is striking that our estimates for the energy and timescale of CSM ejection are consistent with existing PPI models.

It is also important to note that in all non-rotating PPI models the star collapses into a black hole, and none of these models reach the luminosity of SLSNe unless the initial star was rotating rapidly enough to create a magnetar \citep{Woosley2017}. Indeed, \citet{Nicholl2015a} estimated that to power the observed long-lived SLSNe via CSM interaction, densities $\rho_{\rm{CSM}} \gtrsim 10^{-12} \rm{g cm}^{-3}$ and masses comparable to the ejected $M_{\rm ej}$ are probably needed. This is much higher than our inferred density and CSM mass for SN2019szu, and so it is not clear that this event can be solely powered by CSM interaction. PPI SN models reach peak luminosities of $3 \times 10^{43}$\,erg\,s$^{-1}$ from collisions with PPI shells \citep{Sukhbold2016, Woosley2017}. Comparing this to the bolometric light curve of SN2019szu, this extreme would result in only a maximum of a 10\% contribution at peak. If sustained at late times, this would have a more pronounced effect but falls within the error bars of the bolometric luminosity measured.

A more luminous interaction, $\sim{10^{44}}$\,erg\,s$^{-1}$, may be possible if the final SN ejecta interacts with a massive PPI shell (in contrast to the fainter shell-shell collisions). The nebular emission lines from CO material suggest this is unlikely to occur in SN2019szu. In luminous PPI models (Helium cores $\ge 48$\,M$_\odot$, or CO cores $\gtrsim 40$\,M$_\odot$), the bulk of CSM mass is in the first, He-rich pulse, which may or may not be present in SN2019szu depending on whether the He layer was already stripped. The low-mass pulses of CO material come later. The low-density, oxygen-rich CSM producing the emission lines suggests that the CO shells have not been entirely overrun by the SN ejecta, so the SN ejecta cannot have reached any massive He-rich shell even if it was present. Interaction of a massive ejecta with CO shells of $\lesssim 1$\,M$_\odot$ will not contribute a large amount of luminosity, as the fraction of energy thermalised is roughly $M_{\rm ej}/(M_{\rm ej}+M_{\rm CSM})\ll 1$. We do however note that there are lots of uncertainties associated with the final stages of massive star evolution, and so other unknown scenarios involving eruptive mass loss from stripped stars could provide an alternative explanation for SN2019szu, where interaction could contribute a larger fraction of the luminosity. In any case, the oxygen-rich CSM, luminous early plateau, and consistency of the peak luminosity and our estimated ejecta mass with the engine-powered models of \citet{Woosley2017}, make PPI a compelling explanation for the mass-loss in SN2019szu.

\subsection{Implications for other SLSNe and PPI candidates}

The definitive presence of nearby CSM could explain other shared aspects of the SLSN population. The combination of a pre-explosion plateau and the very early appearance of nebular line emission in SN2019szu shows that nebular line emission during the photospheric phase of SLSNe can be an indicator of mass ejection shortly before explosion. The 7300\,\AA\ lines in the spectra of other SLSNe such as SN2007bi, PTF12dam, LSQ14an and SN2015bn may therefore reveal recent mass ejection, potentially driven by the PPI mechanism, in those events too. We note that these are all long-lived SLSNe, likely indicating large ejecta (and therefore progenitor core) masses.

Many SLSNe also show signs of a pre-maximum bump in their light curves \citep{Leloudas2012,Nicholl2015b,Nicholl2016d, Smith2016}, which could potentially be explained by multiple shells of CSM produced by PPI eruptions \citep{Woosley2007}. However, if this was the case we might expect to see spectral evidence of this interaction. Early spectra of other SLSNe such as LSQ14bdq with pre-maximum bumps do not show evidence of broad emission lines  \citep{Nicholl2015b}. Therefore, other explanations such as post-shock cooling of extended stellar material or a recombination wave in the ejecta may be more plausible explanations for some events \citep{Nicholl2015b, Leloudas2012}. However, no spectrum has been obtained during the pre-maximum bump of a SLSN, so it is also possible that spectroscopic signatures of CSM during the bump could be erased by the time of maximum light, e.g.~if the ejecta has overrun the CSM.

Post-maximum bumps in the light curve could also indicate interaction with PPI shells. The long term light curve of SN2017egm shows multiple late time bumps as well as varying levels of decline (Figure \ref{fig:bb comparisons}) \citep{Lin2023, Zhu2023}. \citet{Lin2023} reproduce the evolution of this event using four distinct shells of CSM produced by PPI ejections from a star with a 48–51\,M$_\odot$ He core. This is similar to the proposed He-core mass of SN2019szu and so we might expect to see similar features between the two events. SN2017egm also displays very little [O I] $\lambda$6300 compared to [O II] $\lambda\lambda$7320,7330 which is explained by the high temperatures and therefore ionised neutral oxygen for the event. However, the light curve shapes differ with SN2017egm having a sharper peak, and a faster overall decline. The authors attribute this sharp peak to forward shock between the SN ejecta and the nearest CSM shell. SN2017egm also displays a shorter rise time of $\sim$30 days from explosion to peak \citep{Bose2018}. Another key difference is the environments in which these events occurred. SN2017egm is unique in that it originated from a large galaxy with a high metallicty \citep{Nicholl2017c}. SN2019szu originated in small dwarf galaxy with much lower metallicity, an environment where a rapidly rotating magnetar may be more likely to form. The differences in these events could be explained by a difference in internal engine powering, as an engine is required to match the luminosity of SN2019szu \citep{Woosley2017}, and cannot be explained by CSM interaction alone.

We can see clear observational evidence for circumstellar material in other SLSNe. In iPTF16eh, a Mg II resonance doublet was observed to change from blueshifted emission to redshifted emission over time \citep{Lunnan2018a}. This is explained by reflection of light from a detached shell of CSM surrounding the SN. This material had a velocity of 3300\,\kms\, and was thought to have been ejected 32 years prior to the supernova explosion \citep{Lunnan2018a}, a much longer timescale than derived for SN2019szu. In the case of SN2019szu, we do not see this changing Doppler shift, as the ejecta has already collided with the CSM. However, the mechanism proposed to produce the CSM in iPTF16eh is also pair instability ejections, but in that case from a more massive progenitor with a He core mass of $\sim 51-53$\,M$_{\odot}$, which experiences the PPI earlier before explosion.

Some SLSNe show evidence for interaction only at late times with the appearance of broad H$\alpha$ emission. In SN2018bsz, this feature is multi-component and appears at $\sim$30 days, accompanied by other hydrogen lines. \citet{Pursiainen2022} explain this using highly aspherical CSM with several emitting regions. In iPTF15esb, iPTF16bad and iPTF13ehe, this feature emerged at +73, +97 and +251 days respectively, implying the progenitors lost their hydrogen envelope several decades before the SN explosion, leading to a neutral hydrogen shell \citep{Yan2015, Yan2018}. \citet{Yan2018} suggest this eruptive mass loss could be common in SLSN progenitors. iPTF15esb in particular had a triple-peaked light curve, which could be explained by shells of CSM with a total mass $\sim$ 0.01\,M$_{\odot}$. Collisions between these shells or between shells and ejecta could provide the excess luminosity to power the light curve undulations in iPTF15esb. Undulations during the declining phase in other SLSNe have also been attributed to interaction \citep{Nicholl2016a,Inserra2017,Hosseinzadeh2022}, though central engine flaring or ionisation fronts are an alternative explanation \citep{Metzger2014,Hosseinzadeh2022}.

Other energetic SNe have shown signs of interaction with oxygen rich material, including the unusual Type Ic SN2010mb \citep{Ben-Ami2014}. This event had a slowly declining light curve thought to be the result of interaction with $\sim$3\,M$_{\odot}$ of CSM. This resulted in spectral features such as a blue quasi-continuum and a strong [O I]$\lambda$5577 emission line at late times. However, this emission line had a narrow core and and required densities of $\sim 10^{-14}$\,g\,cm$^{-3}$. This is more dense than the CSM surrounding SN2019szu but could be partially explained by the slower velocity of the CSM in SN2010mb at 800\,\kms. \citet{Ben-Ami2014} also suggest PPI ejections could be the source of this material. Other events that fall into this emerging population of Ic-CSM include SN2022xxf and SN2021ocs, both of which show signs of interaction with H/He poor CSM \citep{Kuncarayakti2023, Kuncarayakti2022}.

Looking at the inferred properties of the progenitor can also provide contextual clues for pair-instability candidates. For example, the type I SN2016iet was estimated to have had a CO core mass of $\sim 55-120\,\rm{M}_{\odot}$ prior to explosion \citep{Gomez2019}. This event was best modelled with interaction with $\sim$35\,M$_{\odot}$ of CSM, ejected within the last decade before explosion. This leads to a mass loss rate of $\sim$7\,M$_{\odot}$\,yr$^{-1}$, much higher than the inferred rate for SN2019szu. These high masses coupled with the low metallicity host galaxy is within the regime of PPI or pair-instability supernovae and is consistent with PPI models by \citet{Woosley2017}.

In summary, the short timescale between explosion and observation of the 7300\,$\Angstrom$ line in SN2019szu supports the theory that the CSM producing this line was ejected by its progenitor very shortly before explosion. This is supported by the precedent set by other SLSNe which have shown evidence for eruptions close to explosion, albeit on a longer range of timescales. The relatively tight constraints on the CSM mass and timing of ejection make SN2019szu one of the strongest candidates for a PPI SN to date, and suggests that some of these events can form the engines required to reach superluminous magnitudes.

\section{Conclusions}
\label{sec:conclusion}

This paper presents extensive optical follow up of the SLSN SN2019szu. This includes spectra taken over nearly 300 days in rest frame, and photometry over 800 days. This event is one of the slowest evolving SLSNe to date with a rise time of  $\sim 80$ days from explosion to peak and an exponential decline time of $\sim$100 days. This event also displayed a pre-explosion plateau at an absolute magnitude $M_{w}\sim-18.7$\,mag lasting 40 days.

SN2019szu displayed a remarkably early forbidden emission line at 16 days before maximum light during its photospheric phase, the earliest we have ever seen such emission lines. Using models of nebular SN spectra from \citet{Jerkstrand2017b}, we were able to not only determine that this line at $\sim 7300\,\Angstrom$ originated from [O II] $\lambda\lambda7320,7330$ but also deduce parameters of the material of origin. We found that SN2019szu had at least $\sim0.25$\,M$_{\odot}$ of H-poor and O-rich material with a density of $\sim10^{-15}$\,g\,cm$^{-3}$.

The spectra of this event also showed a steep continuum in the blue, combined with a relatively flat continuum redwards of $\sim5500\,\Angstrom$. This unique spectral shape was not well fit by a simple blackbody. Instead we showed that this shape could be recreated by combining a hot blackbody spectrum with that of an interacting SN. Combined with the bumpy light curve, sustained blue colours, and the O-rich CSM needed to produce the [O II] line at early times, this provides strong evidence that SN2019szu was interacting with nearby CSM.  In order for the interaction to occur by 16 days before maximum light, it must have been ejected less than 120 days before explosion (assuming a CSM velocity of 1500\,\kms\ based on the blueshift of the emission lines), suggesting that mass ejection is also responsible for the light curve plateau.

We conclude that producing $\sim0.25$\,M$_{\odot}$ of hydrogen-poor CSM close to the time of explosion is not feasible using known mechanisms, such as stellar winds or eruptions from luminous blue variables. Instead we suggest pulsational pair-instability (PPI) ejections are a promising possibility. The PPI mechanism also can explain the lack of H and He in this CSM. PPI models from a stripped $\sim 40$\,M$_\odot$ CO core are consistent with our estimated CSM energetics and ejection timescale, the duration and luminosity of the pre-explosion plateau, and the estimated ejecta mass from the SN light curve. 

The detailed study of SN2019szu introduces a new observational approach that can be used to find signatures of PPI interactions. Early observations of nebular emission lines alongside the characteristic O II absorption lines could be used to probe the structure of SLSNe. Obtaining these observations as soon as possible after explosion could help provide stricter constraints on when and how much mass is ejected during these PPI ejections. Observing pre-explosion activity will also provide more information on the progenitors with spectroscopic observations during this time helping to unravel the composition and velocity of this material. As growing numbers of SLSNe show evidence for interaction with CSM, adopting this approach will also help answer questions about the explosion mechanisms involved. This will be especially useful in future survey telescopes such as the Vera Rubin Observatory, which will be able to detect precursor activity in time for more detailed follow-up.

\section*{Acknowledgements}

We would like to thank Peter Blanchard for help reducing the Binospec spectrum.

AA and MN are supported by the European Research Council (ERC) under the European Union’s Horizon 2020 research and innovation programme (grant agreement No.~948381). MN also acknowledges funding from the UK Space Agency.

This work was funded by ANID, Millennium Science Initiative, ICN12\_009. 

AJ acknowledges support by the European Research Council (ERC) under the European Union’s Horizon 2020 Research and Innovation Programme (ERC
Starting Grant SUPERSPEC, 803189).

TWC thanks the Max Planck Institute for Astrophysics for hosting her as a guest researcher.

LG, MGB, CPG and TEMB acknowledge financial support from the Spanish Ministerio de Ciencia e Innovaci\'on (MCIN), the Agencia Estatal de Investigaci\'on (AEI) 10.13039/501100011033 under the PID2020-115253GA-I00 HOSTFLOWS project, from Centro Superior de Investigaciones Cient\'ificas (CSIC) under the PIE project 20215AT016, and the program Unidad de Excelencia Mar\'ia de Maeztu CEX2020-001058-M.
LG also acknowledges support from the European Social Fund (ESF) "Investing in your future" under the 2019 Ram\'on y Cajal program RYC2019-027683-I.
CPG also acknowledges financial support from the Secretary of Universities and Research (Government of Catalonia) and by the Horizon 2020 Research and Innovation Programme of the European Union under the Marie Sk\l{}odowska-Curie and the Beatriu de Pin\'os 2021 BP 00168 programme.
TEMB also acknowledges support from the European Union Next Generation EU/PRTR funds under the 2021 Juan de la Cierva program FJC2021-047124-I.

GL is supported by a research grant (19054) from VILLUM FONDEN.

SS acknowledges support from the G.R.E.A.T. research environment, funded by {\em Vetenskapsr\aa det}, the Swedish Research Council, project number 2016-06012.

SJS acknowledges funding from STFC Grant ST/X006506/1 and ST/T000198/1.

Based on observations collected at the European Organisation for Astronomical Research in the Southern Hemisphere, Chile, as part of ePESSTO+ (the advanced Public ESO Spectroscopic Survey for Transient Objects Survey). ePESSTO+ observations were obtained under ESO program IDs 1103.D-0328, 106.216C, 108.220C (PI: Inserra).

LCO data have been obtained via OPTCON proposals (IDs: OPTICON 19B-009, 20A/015 and 20B/003). The OPTICON project has received funding from the European Union’s Horizon 2020 research and innovation programme under grant agreement No 730890.

This work has made use of data from the Asteroid Terrestrial-impact Last Alert System (ATLAS) project. ATLAS is primarily funded to search for near earth asteroids through NASA grants NN12AR55G, 80NSSC18K0284, and 80NSSC18K1575; byproducts of the NEO search include images and catalogs from the survey area. The ATLAS science products have been made possible through the contributions of the University of Hawaii Institute for Astronomy, the Queen's University Belfast, the Space Telescope Science Institute, and the South African Astronomical Observatory.

The Pan-STARRS1 Surveys (PS1) have been made possible through contributions of the Institute for Astronomy, the University of Hawaii, the Pan-STARRS Project Office, the Max-Planck Society and its participating institutes, the Max Planck Institute for Astronomy, Heidelberg and the Max Planck Institute for Extraterrestrial Physics, Garching, The Johns Hopkins University, Durham University, the University of Edinburgh, Queen's University Belfast, the Harvard-Smithsonian Center for Astrophysics, the Las Cumbres Observatory Global Telescope Network Incorporated, the National Central University of Taiwan, the Space Telescope Science Institute, the National Aeronautics and Space Administration under Grant No. NNX08AR22G issued through the Planetary Science Division of the NASA Science Mission Directorate, the National Science Foundation under Grant No. AST-1238877, the University of Maryland, and Eotvos Lorand University (ELTE).

Based on observations obtained with the Samuel Oschin 48-inch Telescope at the Palomar Observatory as part of the Zwicky Transient Facility project. ZTF is supported by the National Science Foundation under Grant No. AST-1440341 and a collaboration including Caltech, IPAC, the Weizmann Institute for Science, the Oskar Klein Center at Stockholm University, the University of Maryland, the University of Washington, Deutsches Elektronen-Synchrotron and Humboldt University, Los Alamos National Laboratories, the TANGO Consortium of Taiwan, the University of Wisconsin at Milwaukee, and Lawrence Berkeley National Laboratories. Operations are conducted by COO, IPAC, and UW.

\section*{Data Availability}

All data in this paper will be made publicly available via WISeREP \citep{Yaron2012}.



\bibliographystyle{mnras}
\bibliography{SN2019szu} 



\section*{Affiliations}
\noindent
{\it
$^{1}$Institute for Gravitational Wave Astronomy and School of Physics and Astronomy, University of Birmingham, Birmingham B15 2TT, UK\\
$^{2}$Astrophysics Research Centre, School of Mathematics and Physics, Queen's University Belfast, Belfast BT7 1NN, UK\\
$^{3}$The Oskar Klein Centre, Department of Astronomy, Stockholm University, AlbaNova, SE-10691 Stockholm, Sweden\\
$^{4}$Space Telescope Science Institute, 3700 San Martin Dr, Baltimore, MD 21218, USA\\
$^{5}$Department of Physics, University of Oxford, Denys Wilkinson Building, Keble Road, Oxford OX1 3RH, UK\\
$^{6}$DTU Space, National Space Institute, Technical University of Denmark, Elektrovej 327, 2800 Kgs. Lyngby, Denmark\\
$^{7}$European Southern Observatory, Alonso de C\'ordova 3107, Casilla 19, Santiago, Chile\\
$^{8}$Millennium Institute of Astrophysics MAS, Nuncio Monsenor Sotero Sanz 100, Off. 104, Providencia, Santiago, Chile\\
$^{9}$Center for Astrophysics | Harvard \& Smithsonian, Cambridge, MA 02138, USA\\
$^{10}$Institute for Astronomy, University of Hawaii, 2680 Woodlawn Drive, Honolulu HI 96822\\
$^{11}$Technische Universit{\"a}t M{\"u}nchen, TUM School of Natural Sciences, Physik-Department, James-Franck-Stra{\ss}e 1, 85748 Garching, Germany\\
$^{12}$Institute of Space Sciences (ICE, CSIC), Campus UAB, Carrer de Can Magrans, s/n, E-08193 Barcelona, Spain.\\
$^{13}$Institut d’Estudis Espacials de Catalunya (IEEC), E-08034 Barcelona, Spain.\\
$^{14}$Universitat Aut\`onoma de Barcelona, E-08193 Bellaterra (Barcelona), Spain \\
$^{15}$Astronomical Observatory, University of Warsaw, Al. Ujazdowskie 4, 00-478 Warszawa, Poland
$^{16}$Cardiff Hub for Astrophysics Research and Technology, School of Physics \& Astronomy, Cardiff University, Queens Buildings, The Parade, Cardiff, CF24 3AA, UK\\
$^{17}$Max-Planck-Institut f{\"u}r Astrophysik, Karl-Schwarzschild Stra{\ss}e 1, 85748 Garching, Germany\\
$^{18}$Astrophysics Research Institute, Liverpool John Moores University, IC2, Liverpool Science Park, 146 Brownlow Hill, Liverpool L3 5RF, UK\\
$^{19}$Department of Physics and Astronomy, The Johns Hopkins University, Baltimore, MD 21218, USA, \\
$^{20}$School of Physics, Trinity College Dublin, The University of Dublin, Dublin 2, Ireland\\
$^{21}$Isaac Newton Group (ING), Apt. de correos 321, E-38700, Santa Cruz de La Palma, Canary Islands, Spain\\
}


\appendix

\section{Photometry Data}
\label{sec:photometry data}
\onecolumn

\begin{longtable}{ccccc}
    \caption{Photometric observations of SN2019szu. Upper limits are indicated by a 1 in the upper limit column.} \\
    \label{tab: photometry} \\
        \hline\hline
        MJD & Magnitude & Mag Error &  Filter & Telescope\\
        & (mag) & (mag) & & \\
        \hline
        57670.4 & >22.54 &  &  i & Pan-STARRS \\
        57706.3 & >21.41 &  &  i & Pan-STARRS \\
        57711.4 & >20.51 &  &  i & Pan-STARRS \\
        58064.4 & >22.16 &  &  i & Pan-STARRS \\
        58699.6 & 21.65 & 0.08 &  w & Pan-STARRS \\
        58704.6 & 21.65 & 0.08 &  w & Pan-STARRS \\
        58732.5 & 21.66 & 0.1 &  w & Pan-STARRS \\
        58750.4 & 21.62 & 0.11 &  w & Pan-STARRS \\
        58780.4 & 19.51 & 0.01 &  w & Pan-STARRS \\
        58816.3 & 18.92 & 0.01 &  w & Pan-STARRS \\
        59100.5 & 21.16 & 0.14 &  i & Pan-STARRS \\
        59127.4 & >19.44 &  &  i & Pan-STARRS \\
        59140.3 & >21.39 &  &  w & Pan-STARRS \\
        59156.3 & 21.68 & 0.2 & i & Pan-STARRS \\
        59161.3 & 21.15 & 0.05 & w & Pan-STARRS \\
        59170.3 & 21.43 & 0.14  & w & Pan-STARRS \\
        59193.2 & 21.31 & 0.08  & w & Pan-STARRS \\
        59208.3 & >21.49 &   & i & Pan-STARRS \\
        59436.6 & 22.43 & 0.1  & w & Pan-STARRS \\
        59455.5 & >22.07 &   & i & Pan-STARRS \\
        59464.5 & >22.81 &   & w & Pan-STARRS \\
        59465.4 & 22.44 & 0.17  & w & Pan-STARRS \\
        59485.5 & >22.20 &   & w & Pan-STARRS \\
        59524.3 & >23.08 &   & w & Pan-STARRS \\
        59827.6 & >21.79 &   & w & Pan-STARRS \\
        59842.4 & >22.83 &   & w & Pan-STARRS \\
        59851.4 & >22.83 &   & w & Pan-STARRS \\
        59876.4 & >23.09 &   & w & Pan-STARRS \\
        59895.3 & >21.61 &   & i & Pan-STARRS \\
        59902.3 & >22.73 &   & w & Pan-STARRS \\
        58705.5 & >19.922 &   & c & ATLAS \\
        58709.5 & >16.667 &   & c & ATLAS \\
        58721.5 & >18.659 &   & c & ATLAS \\
        58729.5 & >20.322 &   & c & ATLAS \\
        58753.5 & >20.562 &   & c & ATLAS \\
        58757.6 & >19.594 &   & c & ATLAS \\
        58761.5 & >20.615 &   & c & ATLAS \\
        58762.4 & >20.583 &   & c & ATLAS \\
        58777.4 & 19.508 & 0.085  & c & ATLAS \\
        58781.4 & 19.311 & 0.085  & c & ATLAS \\
        58785.4 & 19.216 & 0.077  & c & ATLAS \\
        58817.3 & 18.792 & 0.05  & c & ATLAS \\
        58841.3 & >19.207 &   & c & ATLAS \\
        58845.3 & 18.721 & 0.048  & c & ATLAS \\
        58869.2 & 18.716 & 0.06  & c & ATLAS \\
        58703.6 & >20.586 &   & o & ATLAS \\
        58717.5 & >20.426 &   & o & ATLAS \\
        58719.5 & >20.578 &   & o & ATLAS \\
        58723.5 & >21.015 &   & o & ATLAS \\
        58733.6 & >19.755 &   & o & ATLAS \\
        58736.5 & >20.07 &   & o & ATLAS \\
        58745.5 & >19.17 &   & o & ATLAS \\
        58746.4 & >16.861 &   & o & ATLAS \\
        58751.4 & >20.893 &   & o & ATLAS \\
        58755.5 & >20.721 &   & o & ATLAS \\
        58763.4 & >19.752 &   & o & ATLAS \\
        58764.4 & >20.177 &   & o & ATLAS \\
        58765.4 & >19.844 &   & o & ATLAS \\
        58771.4 & >19.938 &   & o & ATLAS \\
        58775.4 & >19.167 &   & o & ATLAS \\
        58779.4 & 19.767 & 0.144  & o & ATLAS \\
        58783.4 & 19.419 & 0.099  & o & ATLAS \\
        58787.4 & 19.459 & 0.112  & o & ATLAS \\
        58789.3 & 19.417 & 0.105  & o & ATLAS \\
        58791.3 & 19.212 & 0.127  & o & ATLAS \\
        58797.4 & 18.759 & 0.228  & o & ATLAS \\
        58798.4 & >19.278 &   & o & ATLAS \\
        58799.3 & 19.106 & 0.165  & o & ATLAS \\
        58800.4 & 19.497 & 0.254  & o & ATLAS \\
        58801.4 & 19.592 & 0.256  & o & ATLAS \\
        58802.3 & 18.854 & 0.093  & o & ATLAS \\
        58803.4 & 19.252 & 0.166  & o & ATLAS \\
        58807.4 & 19.161 & 0.091  & o & ATLAS \\
        58811.3 & 19.093 & 0.092  & o & ATLAS \\
        58815.3 & 18.941 & 0.257  & o & ATLAS \\
        58819.3 & 18.903 & 0.141  & o & ATLAS \\
        58821.3 & 19.2 & 0.168  & o & ATLAS \\
        58824.3 & 19.255 & 0.172  & o & ATLAS \\
        58825.3 & 18.877 & 0.162  & o & ATLAS \\
        58826.3 & 19.278 & 0.177  & o & ATLAS \\
        58827.3 & 19.086 & 0.12  & o & ATLAS \\
        58828.3 & 19.028 & 0.119  & o & ATLAS \\
        58829.3 & 18.788 & 0.154  & o & ATLAS \\
        58830.3 & 18.961 & 0.124  & o & ATLAS \\
        58831.3 & >18.27 &   & o & ATLAS \\
        58833.3 & 19.165 & 0.082  & o & ATLAS \\
        58837.3 & >19.343 & 0.1  & o & ATLAS \\
        58839.3 & >17.57 &   & o & ATLAS \\
        58851.3 & 18.921 & 0.259  & o & ATLAS \\
        58852.3 & 19.253 & 0.166  & o & ATLAS \\
        58855.3 & 18.762 & 0.176  & o & ATLAS \\
        58867.3 & 19.051 & 0.099  & o & ATLAS \\
        58871.2 & 19.191 & 0.098  & o & ATLAS \\
        58875.2 & 19.329 & 0.13  & o & ATLAS \\
        58877.2 & >18.136 &   & o & ATLAS \\
        58775.2 & 19.677 & 0.093  & g & ZTF \\
        58778.2 & 19.59 & 0.102  & g & ZTF \\
        58781.2 & 19.365 & 0.104  & g & ZTF \\
        58785.2 & 19.209 & 0.067  & g & ZTF \\
        58789.2 & 19.186 & 0.065  & g & ZTF \\
        58792.2 & 19.011 & 0.085  & g & ZTF \\
        58797.2 & 18.789 & 0.118  & g & ZTF \\
        58800.2 & 18.79 & 0.11  & g & ZTF \\
        58803.2 & 18.63 & 0.096  & g & ZTF \\
        58812.1 & 18.633 & 0.039  & g & ZTF \\
        58833.1 & 18.804 & 0.053 & g & ZTF \\
        58837.1 & 18.754 & 0.054  & g & ZTF \\
        58852.1 & 18.97 & 0.116  & g & ZTF \\
        58856.1 & 18.677 & 0.081  & g & ZTF \\
        58863.1 & 18.767 & 0.056  & g & ZTF \\
        58693.5 & 21.227 & 0.292  & r & ZTF \\
        58773.3 & 20.284 & 0.263  & r & ZTF \\
        58775.2 & 19.732 & 0.105  & r & ZTF \\
        58778.3 & 19.732 & 0.187  & r & ZTF \\
        58781.3 & 19.84 & 0.267  & r & ZTF \\
        58792.1 & 19.181 & 0.105  & r & ZTF \\
        58797.1 & 19.075 & 0.18  & r & ZTF \\
        58800.3 & 18.898 & 0.11  & r & ZTF \\
        58803.2 & 19.053 & 0.121  & r & ZTF \\
        58806.2 & 19.125 & 0.085  & r & ZTF \\
        58812.2 & 18.908 & 0.054  & r & ZTF \\
        58833.1 & 19.074 & 0.131 & r & ZTF \\
        58837.1 & 18.977 & 0.07  & r & ZTF \\
        58847.1 & 18.933 & 0.221  & r & ZTF \\
        58852.2 & 18.965 & 0.118  & r & ZTF \\
        58856.1 & 18.988 & 0.087  & r & ZTF \\
        58860.1 & 19.053 & 0.148  & r & ZTF \\
        58863.1 & 19.033 & 0.094  & r & ZTF \\
        58878.1 & 19.31 & 0.309  & r & ZTF \\
        58808.4 & 18.414 & 0.336  & V & UVOT \\
        58816.3 & 18.639 & 0.356  & V & UVOT \\
        58824.8 & >17.995 &   & V & UVOT \\
        58825.9 & >18.319 &   & V & UVOT \\
        58830.1 & 18.804 & 0.376  & V & UVOT \\
        58840.6 & >18.313 &   & V & UVOT \\
        58844.6 & >17.8 &   & V & UVOT \\
        58847.3 & >17.837 &   & V & UVOT \\
        58852.4 & 18.271 & 0.265  & V & UVOT \\
        58856.4 & >18.094 &   & V & UVOT \\
        58860.3 & >18.154 &   & V & UVOT \\
        58998.3 & >19.076 &   & V & UVOT \\
        59003.8 & >19.184 &   & V & UVOT \\
        59008.8 & >19.14 &   & V & UVOT \\
        59018.6 & >19.27 &    & V & UVOT \\
        59028.9 & >18.704 &    & V & UVOT \\
        59038.9 & >18.528 &    & V & UVOT \\
        58808.4 & 18.874 & 0.25  & B & UVOT \\
        58816.3 & 18.699 & 0.2  & B & UVOT \\
        58824.8 & 18.794 & 0.192  & B & UVOT \\
        58825.8 & 18.601 & 0.16  & B & UVOT \\
        58830.1 & 18.554 & 0.149  & B & UVOT \\
        58840.6 & 19.111 & 0.235  & B & UVOT \\
        58844.6 & 18.817 & 0.258  & B & UVOT \\
        58847.3 & 18.818 & 0.226  & B & UVOT \\
        58852.4 & 19.0 & 0.247  & B & UVOT \\
        58856.4 & 18.978 & 0.294  & B & UVOT \\
        58860.3 & 18.838 & 0.229  & B & UVOT \\
        58998.3 & 19.655 & 0.246  & B & UVOT \\
        59003.8 & 20.433 & 0.33  & B & UVOT \\
        59008.8 & 20.247 & 0.252  & B & UVOT \\
        59018.6 & 20.051 & 0.318  & B & UVOT \\
        59028.9 & >19.679 &    & B & UVOT \\
        59038.9 & >19.577 &    & B & UVOT \\
        58808.4 & 17.854 & 0.164  & U & UVOT \\
        58816.3 & 18.143 & 0.184  & U & UVOT \\
        58820.3 & 17.839 & 0.172  & U & UVOT \\
        58824.8 & 18.074 & 0.156  & U & UVOT \\
        58825.8 & 17.703 & 0.118  & U & UVOT \\
        58830.1 & 17.957 & 0.132  & U & UVOT \\
        58840.6 & 17.75 & 0.116  & U & UVOT \\
        58844.6 & 17.983 & 0.188  & U & UVOT \\
        58847.3 & 17.805 & 0.146  & U & UVOT \\
        58852.4 & 17.846 & 0.14  & U & UVOT \\
        58856.4 & 17.66 & 0.152  & U & UVOT \\
        58860.3 & 17.756 & 0.138  & U & UVOT \\
        58992.7 & 19.205 & 0.146  & U & UVOT \\
        58998.3 & 19.16 & 0.183  & U & UVOT \\
        59003.8 & 19.317 & 0.154  & U & UVOT \\
        59008.8 & 19.114 & 0.112  & U & UVOT \\
        59018.6 & 19.263 & 0.182  & U & UVOT \\
        59028.9 & 19.533 & 0.345  & U & UVOT \\
        59038.8 & >19.31 &   & U & UVOT \\
        58808.4 & 18.959 & 0.243  & UVW1 & UVOT \\
        58816.3 & 18.942 & 0.218  & UVW1 & UVOT \\
        58820.3 & 18.841 & 0.176  & UVW1 & UVOT \\
        58824.8 & 18.566 & 0.161  & UVW1 & UVOT \\
        58825.8 & 18.477 & 0.144  & UVW1 & UVOT \\
        58830.1 & 18.485 & 0.134  & UVW1 & UVOT \\
        58840.6 & 18.471 & 0.132  & UVW1 & UVOT \\
        58844.6 & 18.805 & 0.214  & UVW1 & UVOT \\
        58847.3 & 18.38 & 0.151  & UVW1 & UVOT \\
        58852.4 & 18.217 & 0.12  & UVW1 & UVOT \\
        58856.4 & 18.618 & 0.202  & UVW1 & UVOT \\
        58860.3 & 18.706 & 0.172  & UVW1 & UVOT \\
        59028.9 & >19.442 &   & UVW1 & UVOT \\
        59038.8 & >19.492 &   & UVW1 & UVOT \\
        58808.4 & 19.438 & 0.222  & UVM2 & UVOT \\
        58816.3 & 19.364 & 0.167  & UVM2 & UVOT \\
        58824.8 & 18.997 & 0.149  & UVM2 & UVOT \\
        58825.9 & 19.027 & 0.136  & UVM2 & UVOT \\
        58830.1 & 19.149 & 0.138  & UVM2 & UVOT \\
        58840.6 & 18.797 & 0.115  & UVM2 & UVOT \\
        58844.6 & 19.151 & 0.175  & UVM2 & UVOT \\
        58847.3 & 19.001 & 0.18  & UVM2 & UVOT \\
        58852.4 & 19.14 & 0.139  & UVM2 & UVOT \\
        58856.4 & 18.966 & 0.156  & UVM2 & UVOT \\
        58860.3 & 18.83 & 0.118  & UVM2 & UVOT \\
        58862.4 & 18.901 & 0.167  & UVM2 & UVOT \\
        58870.3 & 18.901 & 0.17  & UVM2 & UVOT \\
        59028.9 & 20.379 & 0.298  & UVM2 & UVOT \\
        59038.9 & 20.105 & 0.228  & UVM2 & UVOT \\
        58808.4 & 19.598 & 0.257  & UVW2 & UVOT \\
        58816.3 & 19.623 & 0.216  & UVW2 & UVOT \\
        58824.8 & 19.058 & 0.158  & UVW2 & UVOT \\
        58825.9 & 19.373 & 0.172  & UVW2 & UVOT \\
        58830.1 & 19.341 & 0.161  & UVW2 & UVOT \\
        58840.6 & 19.424 & 0.165  & UVW2 & UVOT \\
        58844.6 & 19.396 & 0.206  & UVW2 & UVOT \\
        58847.3 & 19.064 & 0.168  & UVW2 & UVOT \\
        58852.4 & 19.376 & 0.167  & UVW2 & UVOT \\
        58856.4 & 18.934 & 0.166  & UVW2 & UVOT \\
        58860.3 & 19.254 & 0.163  & UVW2 & UVOT \\
        59028.9 & 20.2 & 0.265  & UVW2 & UVOT \\
        59038.9 & >19.904 &   & UVW2 & UVOT \\
        58805.0 & 18.61 & 0.09  & g & LCO \\
        59078.0 & >20.72 &   & g & LCO \\
        59130.9 & 20.93 & 0.12  & g & LCO \\
        59058.1 & 20.76 & 0.1  & g & LCO \\
        59083.1 & 20.93 & 0.12  & g & LCO \\
        58798.9 & 18.76 & 0.06  & g & LCO \\
        59072.0 & >20.84 &   & g & LCO \\
        58825.8 & 18.49 & 0.09  & g & LCO \\
        59169.1 & 21.18 & 0.06  & g & NTT \\
        58851.8 & 18.67 & 0.06  & g & LCO \\
        58846.1 & 18.76 & 0.1  & g & LCO \\
        59152.2 & 21.38 & 0.1  & g & NTT \\
        58843.8 & 18.54 & 0.07  & g & LCO \\
        58858.0 & 18.7 & 0.05  & g & LCO \\
        59115.5 & 20.72 & 0.11  & g & LCO \\
        59046.1 & 20.49 & 0.1  & g & LCO \\
        59103.0 & >19.84 &   & g & LCO \\
        59063.6 & 20.98 & 0.23  & g & LCO \\
        59439.2 & 22.37 & 0.09  & g & NTT \\
        59031.0 & 20.33 & 0.13  & g & LCO \\
        58837.8 & 18.47 & 0.06  & g & LCO \\
        58814.2 & 18.59 & 0.08  & g & LCO \\
        59089.5 & >20.51 &   & g & LCO \\
        59198.1 & 21.6 & 0.06  & g & NTT \\
        58832.1 & 18.57 & 0.08  & g & LCO \\
        59109.6 & 20.68 & 0.09  & g & LCO \\
        59136.6 & 20.89 & 0.1  & g & LCO \\
        59052.1 & 20.57 & 0.09  & g & LCO \\
        58820.0 & 18.58 & 0.09  & g & LCO \\
        59124.7 & >20.71 &   & g & LCO \\
        59058.2 & 20.89 & 0.08  & r & LCO \\
        59130.9 & 21.01 & 0.09  & r & LCO \\
        59078.0 & >20.56 &   & r & LCO \\
        58805.0 & 19.0 & 0.04  & r & LCO \\
        59083.1 & 20.82 & 0.11  & r & LCO \\
        59072.0 & >20.59 &   & r & LCO \\
        58798.9 & 19.04 & 0.04  & r & LCO \\
        58851.8 & 18.97 & 0.05  & r & LCO \\
        58825.8 & 19.0 & 0.12  & r & LCO \\
        59146.0 & 21.74 & 0.1  & r & NTT \\
        58858.0 & 18.99 & 0.08  & r & LCO \\
        58846.1 & 19.05 & 0.06  & r & LCO \\
        59439.2 & >23.2 &   & r & NTT \\
        59063.6 & >21.41 &   & r & LCO \\
        59046.1 & 20.86 & 0.13  & r & LCO \\
        59103.0 & >19.7 &   & r & LCO \\
        59115.5 & 20.88 & 0.13  & r & LCO \\
        59198.1 & 22.2 & 0.11  & r & NTT \\
        58837.8 & 18.89 & 0.09  & r & LCO \\
        59089.5 & >20.72 &   & r & LCO \\
        58814.2 & 18.89 & 0.07  & r & LCO \\
        59031.0 & >20.96 &   & r & LCO \\
        58820.0 & 18.86 & 0.08  & r & LCO \\
        59124.7 & >20.7 &   & r & LCO \\
        58832.1 & 18.9 & 0.08  & r & LCO \\
        59109.6 & 20.67 & 0.08  & r & LCO \\
        59136.6 & 20.74 & 0.11  & r & LCO \\
        58825.8 & >19.42 &   & i & LCO \\
        58851.8 & 19.38 & 0.07  & i & LCO \\
        59072.0 & >20.59 &   & i & LCO \\
        58798.9 & 19.59 & 0.07  & i & LCO \\
        59198.1 & 21.99 & 0.11  & i & NTT \\
        59083.1 & >21.19 &   & i & LCO \\
        59152.2 & 21.89 & 0.14  & i & NTT \\
        59439.2 & >23.01 &   & i & NTT \\
        59130.9 & >20.97 &   & i & LCO \\
        59058.2 & >21.36 &   & i & LCO \\
        58805.0 & 19.43 & 0.05  & i & LCO \\
        59078.0 & >20.97 &   & i & LCO \\
        59052.1 & >21.39 &   & i & LCO \\
        58820.0 & 19.46 & 0.08  & i & LCO \\
        59124.7 & >20.69 &   & i & LCO \\
        59109.6 & 21.48 & 0.19  & i & LCO \\
        58832.1 & 19.31 & 0.05  & i & LCO \\
        59136.7 & >21.21 &   & i & LCO \\
        59169.2 & 21.53 & 0.1  & i & NTT \\
        59031.0 & >20.69 &   & i & LCO \\
        58837.8 & 19.41 & 0.05  & i & LCO \\
        58814.2 & 19.43 & 0.04  & i & LCO \\
        59089.5 & >20.53 &   & i & LCO \\
        59046.1 & >20.98 &   & i & LCO \\
        59103.0 & >19.66 &   & i & LCO \\
        59063.6 & >21.12 &   & i & LCO \\
        59115.5 & >21.11 &   & i & LCO \\
        58858.0 & 19.5 & 0.08  & i & LCO \\
        58846.1 & 19.41 & 0.11  & i & LCO \\
        58805.0 & 18.61 & 0.09  & g & LCO \\
        59078.0 & >20.72 &   & g & LCO \\
        59130.9 & 20.93 & 0.12  & g & LCO \\
        59058.1 & 20.76 & 0.1  & g & LCO \\
        59083.1 & 20.93 & 0.12  & g & LCO \\
        58798.9 & 18.76 & 0.06  & g & LCO \\
        59072.0 & >20.84 &   & g & LCO \\
        58825.8 & 18.49 & 0.09  & g & LCO \\
        59169.1 & 21.18 & 0.06  & g & NTT \\
        58851.8 & 18.67 & 0.06  & g & LCO \\
        58846.1 & 18.76 & 0.1  & g & LCO \\
        59152.2 & 21.38 & 0.1  & g & NTT \\
        58843.8 & 18.54 & 0.07  & g & LCO \\
        58858.0 & 18.7 & 0.05  & g & LCO \\
        59115.5 & 20.72 & 0.11  & g & LCO \\
        59046.1 & 20.49 & 0.1  & g & LCO \\
        59103.0 & >19.84 &   & g & LCO \\
        59063.6 & 20.98 & 0.23  & g & LCO \\
        59439.2 & 22.37 & 0.09  & g & NTT \\
        59031.0 & 20.33 & 0.13  & g & LCO \\
        58837.8 & 18.47 & 0.06  & g & LCO \\
        58814.2 & 18.59 & 0.08  & g & LCO \\
        59089.5 & >20.51 &   & g & LCO \\
        59198.1 & 21.6 & 0.06  & g & NTT \\
        58832.1 & 18.57 & 0.08  & g & LCO \\
        59109.6 & 20.68 & 0.09  & g & LCO \\
        59136.6 & 20.89 & 0.1  & g & LCO \\
        59052.1 & 20.57 & 0.09  & g & LCO \\
        58820.0 & 18.58 & 0.09  & g & LCO \\
        59124.7 & >20.71 &   & g & LCO \\
        59058.2 & 20.89 & 0.08  & r & LCO \\
        59130.9 & 21.01 & 0.09  & r & LCO \\
        59078.0 & >20.56 &   & r & LCO \\
        58805.0 & 19.0 & 0.04  & r & LCO \\
        59083.1 & 20.82 & 0.11  & r & LCO \\
        59072.0 & >20.59 &   & r & LCO \\
        58798.9 & 19.04 & 0.04  & r & LCO \\
        58851.8 & 18.97 & 0.05  & r & LCO \\
        58825.8 & 19.0 & 0.12  & r & LCO \\
        59146.0 & 21.74 & 0.1  & r & NTT \\
        58858.0 & 18.99 & 0.08  & r & LCO \\
        58846.1 & 19.05 & 0.06  & r & LCO \\
        59439.2 & >23.2 &   & r & NTT \\
        59063.6 & >21.41 &   & r & LCO \\
        59046.1 & 20.86 & 0.13  & r & LCO \\
        59103.0 & >19.7 &   & r & LCO \\
        59115.5 & 20.88 & 0.13  & r & LCO \\
        59198.1 & 22.2 & 0.11  & r & NTT \\
        58837.8 & 18.89 & 0.09  & r & LCO \\
        59089.5 & >20.72 &   & r & LCO \\
        58814.2 & 18.89 & 0.07  & r & LCO \\
        59031.0 & >20.96 &   & r & LCO \\
        58820.0 & 18.86 & 0.08  & r & LCO \\
        59124.7 & >20.7 &   & r & LCO \\
        58832.1 & 18.9 & 0.08  & r & LCO \\
        59109.6 & 20.67 & 0.08  & r & LCO \\
        59136.6 & 20.74 & 0.11  & r & LCO \\
        58825.8 & >19.42 &   & i & LCO \\
        58851.8 & 19.38 & 0.07  & i & LCO \\
        59072.0 & >20.59 &   & i & LCO \\
        58798.9 & 19.59 & 0.07  & i & LCO \\
        59198.1 & 21.99 & 0.11  & i & NTT \\
        59083.1 & >21.19 &   & i & LCO \\
        59152.2 & 21.89 & 0.14  & i & NTT \\
        59439.2 & >23.01 &   & i & NTT \\
        59130.9 & >20.97 &   & i & LCO \\
        59058.2 & >21.36 &   & i & LCO \\
        58805.0 & 19.43 & 0.05  & i & LCO \\
        59078.0 & >20.97 &   & i & LCO \\
        59052.1 & >21.39 &   & i & LCO \\
        58820.0 & 19.46 & 0.08  & i & LCO \\
        59124.7 & >20.69 &   & i & LCO \\
        59109.6 & 21.48 & 0.19  & i & LCO \\
        58832.1 & 19.31 & 0.05  & i & LCO \\
        59136.7 & >21.21 &   & i & LCO \\
        59169.2 & 21.53 & 0.1  & i & NTT \\
        59031.0 & >20.69 &   & i & LCO \\
        58837.8 & 19.41 & 0.05  & i & LCO \\
        58814.2 & 19.43 & 0.04  & i & LCO \\
        59089.5 & >20.53 &   & i & LCO \\
        59046.1 & >20.98 &   & i & LCO \\
        59103.0 & >19.66 &   & i & LCO \\
        59063.6 & >21.12 &   & i & LCO \\
        59115.5 & >21.11 &   & i & LCO \\
        58858.0 & 19.5 & 0.08  & i & LCO \\
        58846.1 & 19.41 & 0.11  & i & LCO \\
        \hline \hline 
\end{longtable}


\bsp	
\label{lastpage}
\end{document}